\documentclass[11pt,a4paper]{article}
\usepackage[T1]{fontenc}
\usepackage[utf8]{inputenc}
\usepackage{lmodern}
\usepackage[a4paper,left=28mm,right=28mm,top=28mm,bottom=28mm]{geometry}
\usepackage{amsmath,amssymb,amsthm,mathtools}
\usepackage{microtype}
\usepackage{xcolor}
\usepackage{tikz}
\usetikzlibrary{arrows.meta,positioning,calc,fit,decorations.pathreplacing}
\usepackage[font=small,labelfont=bf]{caption}
\usepackage{enumitem}
\usepackage{cite}
\usepackage[hidelinks]{hyperref}
\hypersetup{pdftitle={Quantum State Preparation without Disturbing the Memory},pdfauthor={},pdfsubject={Non-disturbing quantum preparation, maximal divergences, and hypothesis testing}}
\newtheorem{theorem}{Theorem}[section]
\newtheorem{proposition}[theorem]{Proposition}
\newtheorem{lemma}[theorem]{Lemma}

\theoremstyle{definition}
\newtheorem{definition}[theorem]{Definition}
\newtheorem*{definition*}{Definition}
\theoremstyle{remark}
\newtheorem{remark}[theorem]{Remark}
\newtheorem*{remark*}{Remark}
\DeclareMathOperator{\tr}{tr}
\DeclareMathOperator{\supp}{supp}
\DeclareMathOperator{\diag}{diag}
\DeclareMathOperator{\id}{id}
\DeclareMathOperator{\co}{co}

\newcommand{\BS}{D_{\mathrm{BS}}}
\newcommand{\KL}{D_{\mathrm{KL}}}
\newcommand{\D}{\mathcal D}
\newcommand{\cE}{\mathcal E}
\newcommand{\cL}{\mathcal L}
\newcommand{\ket}[1]{\lvert #1\rangle}
\newcommand{\bra}[1]{\langle #1\rvert}
\newcommand{\norm}[1]{\lVert #1\rVert}

\newcommand{\doi}[1]{\href{https://doi.org/#1}{\nolinkurl{doi:#1}}}
\newcommand{\arxiv}[1]{\href{https://arxiv.org/abs/#1}{\nolinkurl{arXiv:#1}}}
\title{Quantum Preparation without Disturbance: Information Costs and Hypothesis Testing}
\author{Pablo Costa Rico$^{1}$,  Robert Salzmann$^{1}$, Mario Berta$^{1}$, Ángela Capel$^{2,3}$ \\[0.8em]
 \parbox{0.92\textwidth}{\centering\small
 $^{1}$Institute for Quantum Information, RWTH Aachen University,
 Aachen, Germany\\
 $^{2}$Department of Applied Mathematics and Theoretical Physics, University of Cambridge,  Cambridge CB3 0WA, United Kingdom \\
 $^{3}$Fachbereich Mathematik, Universit\"at T\"ubingen, 72076 T\"ubingen, Germany}}
\date{}

\newcommand{\cR}{\mathcal{R}}

\begin{document}
\maketitle
\vspace{-1.5em}
\begin{abstract}
We determine the minimum information cost of preparing either of two quantum
states with one common device while preserving its memory. Measured by the
Umegaki relative entropy between the two possible memory states, this minimum
is the Belavkin--Staszewski relative entropy of the target states. If the
memory may instead be consumed, the minimum is their Umegaki relative entropy.
More generally, for any divergence satisfying data processing, memory
preservation selects the maximal extension of its classical restriction.
We also study the cost of requiring only finitely many outputs, with the costs increasing from Umegaki to Belavkin--Staszewski
relative entropy as the number of outputs grows. Finally, we consider
hypothesis testing of an unknown device over independent runs. The optimal Stein exponent per run converges to the
Belavkin--Staszewski relative entropy whenever the number of outputs per run
diverges. This gives an operational characterization of the
Belavkin--Staszewski relative entropy in quantum statistics.
\end{abstract}

\tableofcontents

\section{Introduction}\label{sec:introduction}

State preparation provides a natural setting for studying how a physical
device uses information. Besides specifying the desired output, one may
consider the record that guides the preparation and whether it remains
available afterwards. This distinction is particularly relevant for quantum
records, which can be disturbed when their information is used.

We therefore ask how much information a memory must contain to prepare
either of two quantum states using one common device. The device receives
a memory state $\omega_i$ and must produce $\rho_i$, for $i=0,1$.
We measure the information cost by the \emph{Umegaki relative entropy}
$D(\omega_0\Vert\omega_1)$, a standard measure of quantum
distinguishability, and minimize over all compatible memories and devices.
If the memory may be consumed, the device can simply transfer its state
to the output, giving the minimum $D(\rho_0\Vert\rho_1)$.

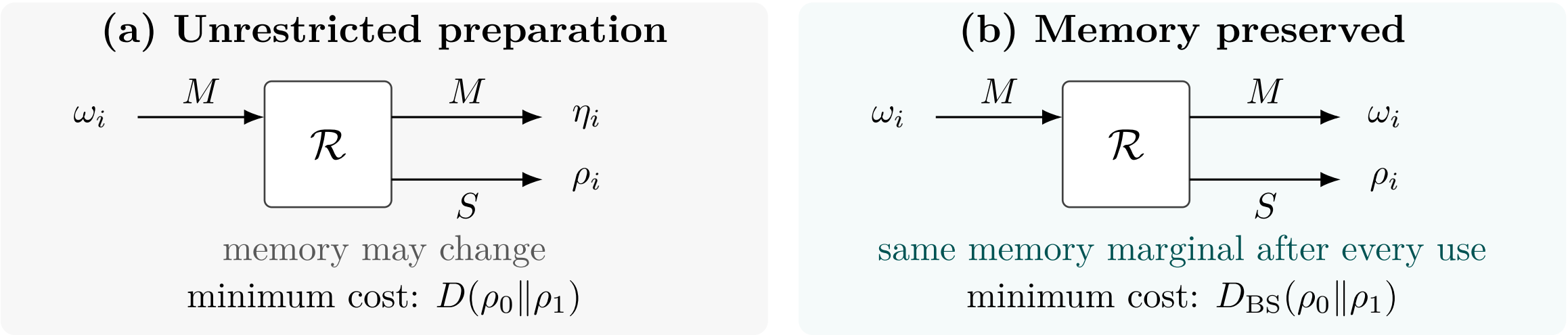
\begin{figure}[htbp]
\centering
\begin{tikzpicture}[>=Latex,x=1cm,y=1cm,font=\small,
  panel/.style={rounded corners=4pt,draw=none},
  device/.style={draw=black!75,fill=white,rounded corners=2pt,line width=.5pt},
  wire/.style={->,line width=.6pt}]
  \path[panel,fill=black!3] (0,0) rectangle (7.5,3.25);
  \node[font=\bfseries] at (3.75,2.96) {(a) Unrestricted preparation};
  \path[device] (2.58,1.25) rectangle (3.82,2.48);
  \node[font=\large] at (3.20,1.865) {$\mathcal R$};
  \node at (.88,2.13) {$\omega_i$};
  \draw[wire] (1.34,2.13)--(2.58,2.13);
  \node[above] at (1.96,2.13) {$M$};
  \draw[wire] (3.82,2.13)--(5.30,2.13);
  \node[above] at (4.56,2.13) {$M$};
  \node at (5.73,2.13) {$\eta_i$};
  \draw[wire] (3.82,1.52)--(5.30,1.52);
  \node[below] at (4.56,1.52) {$S$};
  \node at (5.73,1.52) {$\rho_i$};
  \node[text=black!65] at (3.75,.82) {memory may change};
  \node at (3.75,.36) {minimum cost: $D(\rho_0\Vert\rho_1)$};
  \begin{scope}[xshift=7.8cm]
    \path[panel,fill=teal!4] (0,0) rectangle (7.5,3.25);
    \node[font=\bfseries] at (3.75,2.96) {(b) Memory preserved};
    \path[device] (2.58,1.25) rectangle (3.82,2.48);
    \node[font=\large] at (3.20,1.865) {$\mathcal R$};
    \node at (.88,2.13) {$\omega_i$};
    \draw[wire] (1.34,2.13)--(2.58,2.13);
    \node[above] at (1.96,2.13) {$M$};
    \draw[wire] (3.82,2.13)--(5.30,2.13);
    \node[above] at (4.56,2.13) {$M$};
    \node at (5.73,2.13) {$\omega_i$};
    \draw[wire] (3.82,1.52)--(5.30,1.52);
    \node[below] at (4.56,1.52) {$S$};
    \node at (5.73,1.52) {$\rho_i$};
    \node[text=teal!65!black] at (3.75,.82)
      {same memory marginal after every use};
    \node at (3.75,.36) {minimum cost: $D_{\mathrm{BS}}(\rho_0\Vert\rho_1)$};
  \end{scope}
\end{tikzpicture}
\caption{Preparation with and without memory preservation. The same channel
$\mathcal R$ prepares $\rho_i$ from $\omega_i$ for $i=0,1$.
The displayed minimum costs use Umegaki relative entropy to measure the
initial memory distinguishability and assume
$\supp\rho_0\subseteq\supp\rho_1$.
Outgoing states denote marginals of the joint memory--system state;
correlations are allowed.}
\label{fig:memory}
\end{figure}

We show that preserving the memory changes this minimum to the
\emph{Belavkin--Staszewski (BS) relative entropy}
$\BS(\rho_0\Vert\rho_1)$, another quantum extension of classical
relative entropy that upper bounds $D$; see Figure~\ref{fig:memory}.
Here only the memory marginal is preserved:
correlations with the output are allowed, so the memory can be reused
without producing independent outputs.

A classical label can be retained after preparing the output, making this
preparation naturally non-disturbing. Our result shows that allowing quantum
memories cannot lower this cost under memory preservation: every admissible
preparation can be replaced, at no greater information cost, by a classical
record.

\subsection{The non-commutative preparation picture}

Let $\rho_0,\rho_1\in\D(S)$ be two prescribed output states. A preparation
device consists of a finite-dimensional memory $M$, two initial states
$\omega_0,\omega_1\in\D(M)$, and a common quantum channel
$\cR:M\to M\otimes S$ satisfying
\begin{equation}\label{eq:target}
  \tr_M\cR(\omega_i)=\rho_i,\qquad i=0,1.
\end{equation}

Writing $\mathcal T:=\tr_S\circ\cR$ for the reduced memory evolution,
we call the preparation \emph{stationary} if
\begin{equation}\label{eq:stationarity-constraint}
  \mathcal T(\omega_i)=\omega_i,\qquad i=0,1.
\end{equation}
Repeated use then preserves the memory marginal $\omega_i$ and produces
the marginal $\rho_i$ at every output, with correlations allowed.
The memory dimension is not fixed, and $\omega_0,\omega_1$ need not be
faithful or satisfy support inclusion.
This condition connects preparation with information extraction without
disturbance: Koashi and Imoto \cite{KoashiImoto2002} characterized the
structure of operations that preserve a family of quantum states.

Following Gour and Tomamichel \cite{GourTomamichel2020}, a \emph{quantum
divergence} is a function $\mathbb D$ on pairs of states on finite-dimensional
systems, with values in $[0,+\infty]$, satisfying the data-processing inequality
\begin{equation}
  \mathbb{D}(\Phi(\rho)\|\Phi(\sigma))\le\mathbb{D}(\rho\|\sigma)
  \label{eq:dpi}
\end{equation}
for every quantum channel $\Phi$.
No additivity, continuity, or normalization on equal states is assumed.
Its classical restriction is
\begin{equation}
\mathbb{D}_{\mathrm{cl}}(p\|q)  \coloneqq \mathbb{D}(\diag p\|\diag q)
  \label{eq:classical-restriction}
\end{equation}
for probability vectors $p,q$ on a common finite alphabet. We measure the
\emph{information cost} of the preparation by $\mathbb D(\omega_0\|\omega_1)$.

\begin{theorem}[Stationary Preparation]
\label{thm:preparation}
Let $
\rho_0,\rho_1\in\mathcal D(S)$ satisfy $
\operatorname{supp}\rho_0\subseteq\operatorname{supp}\rho_1
$, and let $\mathbb{D}$ be a quantum divergence, with classical restriction $\mathbb{D}_{\mathrm{cl}}$. Then
\begin{equation}
  \inf_{\substack{M,\omega_0,\omega_1,\cR:M\to M\otimes S\\
                  \tr_M\cR(\omega_i)=\rho_i\ (i=0,1)}}
       \mathbb{D}(\omega_0\|\omega_1)
  =\mathbb{D}(\rho_0\|\rho_1).
  \label{eq:unrestricted-minimum}
\end{equation}
With the additional memory-preservation conditions,
\begin{equation}
  \inf_{\substack{M,\omega_0,\omega_1,\cR:M\to M\otimes S\\
                  \tr_S\cR(\omega_i)=\omega_i\\
                  \tr_M\cR(\omega_i)=\rho_i\ (i=0,1)}}
       \mathbb{D}(\omega_0\|\omega_1)
  =\inf_{\substack{p_0,p_1,\{\tau_z\}\\
                    \sum_zp_i(z)\tau_z=\rho_i\ (i=0,1)}}
       \mathbb D_{\mathrm{cl}}(p_0\Vert p_1).
  \label{eq:stationary-minimum}
\end{equation}
Here $p_0,p_1$ are distributions on a common finite alphabet and $\tau_z$ are states on $S$.
The infimum in \eqref{eq:unrestricted-minimum} is attained. In
\eqref{eq:stationary-minimum}, restricting to classical memories does not
change the infimum. If the classical infimum on the right is attained, the corresponding classical memory
attains \eqref{eq:stationary-minimum}.
\end{theorem}

\begin{proof}
The proof can be found in Theorem \ref{AP-thm:preparation}.
\end{proof}

The stationary preparation identity also admits a robust version. We now
allow small errors in both the retained-memory marginal and the prescribed
output, while keeping the same channel for the two initial states. For
$\varepsilon_{\mathrm s},\varepsilon_{\mathrm o}\geq0$, define
\begin{equation}\label{eq:approximate-stationary-cost}
 I_{\varepsilon_{\mathrm s},\varepsilon_{\mathrm o}}^{\mathbb D}
 (\rho_0\Vert\rho_1)
 :=\inf_{\substack{
 M,\omega_0,\omega_1,\cR:M\to M\otimes S\\
 \|\tr_S\cR(\omega_i)-\omega_i\|_1\leq\varepsilon_{\mathrm s}\\
 \|\tr_M\cR(\omega_i)-\rho_i\|_1\leq\varepsilon_{\mathrm o},\ i=0,1}}
 \mathbb D(\omega_0\Vert\omega_1).
\end{equation}
The infimum is over finite-dimensional memories of unrestricted dimension,
initial memory states, and quantum channels.

\begin{theorem}[Robustness of stationary preparation]
\label{thm:stationary-robustness}
Let $\rho_0,\rho_1>0$ be states on $S$, and let $\mathbb D$ satisfy
data processing with classical restriction
$\mathbb D_{\mathrm{cl}}=D_{\mathrm{KL}}$. Write $d=\dim S$ and set
\[
 \lambda=\min\{\lambda_{\min}(\rho_0),\lambda_{\min}(\rho_1)\},
 \qquad
 L_{\rho_0,\rho_1}
 =\log\frac{2}{\lambda}+\frac{2\sqrt{2}}{\lambda^{5/2}}.
\]
For $\varepsilon_{\mathrm s},\varepsilon_{\mathrm o}\geq0$ satisfying
$\varepsilon_{\mathrm o}+2d\sqrt{\varepsilon_{\mathrm s}}
\leq\lambda/2$, the cost in
\eqref{eq:approximate-stationary-cost} obeys
\begin{equation}\label{eq:stationary-robust-bound}
 0\leq \BS(\rho_0\Vert\rho_1)
 -I_{\varepsilon_{\mathrm s},\varepsilon_{\mathrm o}}^{\mathbb D}
 (\rho_0\Vert\rho_1)\\
 \leq 2L_{\rho_0,\rho_1}
 \bigl(\varepsilon_{\mathrm o}
       +2d\sqrt{\varepsilon_{\mathrm s}}\bigr).
\end{equation}
In particular,
\begin{equation}\label{eq:stationary-robust-limit}
 \lim_{
 \varepsilon_{\mathrm s},\varepsilon_{\mathrm o}\to0^+}
 I_{\varepsilon_{\mathrm s},\varepsilon_{\mathrm o}}^{\mathbb D}
 (\rho_0\Vert\rho_1)
 =\BS(\rho_0\Vert\rho_1).
\end{equation}
\end{theorem}

\begin{proof}
The proof is given in Theorem~\ref{AP-thm:stationary-robustness}.
\end{proof}

\subsection{Maximal extensions and relative entropy}

The classical optimization in Theorem~\ref{thm:preparation} is the
\emph{maximal extension} introduced by Gour and Tomamichel
\cite{GourTomamichel2020},
\begin{equation}
  \widehat{\mathbb{D}}(\rho_0\|\rho_1)
  :=\inf_{\substack{p_0,p_1,\{\tau_z\}\\
                    \sum_z p_i(z)\tau_z=\rho_i\ (i=0,1)}}
       \mathbb{D}_{\mathrm{cl}}(p_0\|p_1).
  \label{eq:maximal-extension}
\end{equation}
where $p_0,p_1$ are probability vectors on a common finite alphabet and
$\tau_z\in\D(S)$. This quantity depends only on
$\mathbb D_{\mathrm{cl}}$. Theorem \ref{thm:preparation} identifies it with the minimum
information cost of memory-preserving preparation.

For the Umegaki relative entropy
$D(\rho\Vert\sigma)=\tr[\rho(\log\rho-\log\sigma)]$ \cite{Umegaki1962},
the classical restriction is the Kullback--Leibler divergence
$\KL(p\Vert q)=\sum_z p(z)\log[p(z)/q(z)]$.
Matsumoto's reverse-test characterization \cite{Matsumoto2010a} identifies
its maximal extension with the Belavkin--Staszewski (BS) relative entropy
\cite{BelavkinStaszewski1982}. For faithful states, it takes the form
\begin{equation}\label{eq:bs-definition-intro}
 \BS(\rho\Vert\sigma)
 =\tr\!\left[\rho\log\!\left(\rho^{1/2}\sigma^{-1}\rho^{1/2}\right)\right].
\end{equation}
The BS relative entropy satisfies data processing, upper bounds $D$, and
coincides with it for commuting states \cite{Matsumoto2010a,HiaiMosonyi2017}.
Thus Theorem~\ref{thm:preparation} gives the two costs in
Figure~\ref{fig:memory}: $D$ for unrestricted preparation and $\BS$ when
the memory is preserved.

\subsection{Overview of further results}

\paragraph{Equivalent preparation pictures.}
The same minimum information cost is obtained by restricting the preparation
channel to be entanglement breaking (Proposition~\ref{prop:EB}), or by
requiring the prescribed output at every step of an infinite sequence while
allowing the memory to evolve (Proposition~\ref{prop:output-stationarity}).

\paragraph{Finite reuse.}
Finite reuse requirements connect unrestricted and stationary preparation. We introduce
$I_n^{\mathbb D}(\rho\Vert\sigma)$, the minimum initial record divergence
needed to produce $n$ emissions with the prescribed marginals, allowing
the memory to evolve between uses. For faithful states and
$\mathbb D=D$, we prove that these minima increase monotonically from
the Umegaki relative entropy to the BS relative entropy:
\begin{equation}\label{eq:interpolation-intro}
 \begin{gathered}
 D(\rho\Vert\sigma)=I_1^D(\rho\Vert\sigma)
 \le I_n^D(\rho\Vert\sigma)\le I_{n+1}^D(\rho\Vert\sigma),\\
 \lim_{n\to\infty}I_n^D(\rho\Vert\sigma)
 =I_\infty^D(\rho\Vert\sigma)=\BS(\rho\Vert\sigma).
 \end{gathered}
\end{equation}
The number of required emissions therefore supplies an operational
interpolation between Umegaki and BS. Each member satisfies data processing,
and the hierarchy collapses for commuting states. For noncommuting targets with support inclusion,
the Umegaki hierarchy is strictly increasing and every finite term is below BS
(Proposition~\ref{prop:strict-hierarchy-general}). For faithful states, both the preparation costs and the universal testing exponents converge to the BS relative entropy at rate $O(n^{-1})$. More generally, every initial divergence
$\mathbb D$ satisfying data processing and reducing classically to
Kullback--Leibler yields the same BS endpoint.

The interpolation in \eqref{eq:interpolation-intro} is in fact
a particular instance of a more general result. For any quantum
divergence $\mathbb D$ satisfying data processing, the corresponding
finite-use preparation costs form a monotone hierarchy between
$\mathbb D$ and its maximal extension $\widehat{\mathbb D}$,
with $I_1^{\mathbb D}=\mathbb D$ and
$I_\infty^{\mathbb D}=\widehat{\mathbb D}$.
If, in addition, $\widehat{\mathbb D}$ is jointly lower
semicontinuous at the target pair, as specified in
Theorem~\ref{thm:hierarchy}, the finite-use costs converge
to this maximal extension as $n\to\infty$.

\paragraph{Hypothesis testing.}
The minimum memory information cost identified in
Theorem~\ref{thm:preparation} also has a hypothesis-testing
interpretation. Suppose a tester receives independently initialized
copies of the memory of a known stationary preparation and aims to
distinguish its two possible memory states. We show in
Theorem~\ref{thm:source-Stein} that, after optimizing the test
for each known preparation and taking the worst case over all
compatible preparations, the asymptotic type-II error exponent
per memory copy is given by the BS relative entropy.

We then consider the setting in which the preparation device is
unknown and only its emitted systems are accessible.
In ordinary quantum Stein testing, we distinguish
independent repetitions of a known pair of states, and their Umegaki
relative entropy determines the error exponent \cite{HiaiPetz1991,OgawaNagaoka2000}.
Here the same unknown device is independently initialized in each run and
emits $n$ systems, all with marginal $\rho$ under the null hypothesis or
$\sigma$ under the alternative. Their joint state is unknown, and
correlations within a run are unrestricted. We require a single test to
work uniformly over all compatible devices. For $n=1$, we recover ordinary
Stein testing with exponent $D(\rho\Vert\sigma)$. For faithful targets
and any fixed type-I error bound $0<\varepsilon<1$, we show that increasing
the number of emissions per run yields the BS exponent:
\begin{equation}\label{eq:main-intro}
 \lim_{m\to\infty}-\frac1m\log\beta_{n_m,m}^{\varepsilon}
 =\BS(\rho\Vert\sigma)
 \qquad\text{whenever }n_m\to\infty.
\end{equation}
Here $\beta_{n,m}^{\varepsilon}$ is the optimal worst-case type-II error
after $m$ independent initializations, so the exponent is per run of
$n_m$ emissions.
No restriction is imposed on the relative growth of $n_m$ and $m$.
This gives a Stein characterization of the BS-entropy using only the emitted systems
and knowledge of their local states.

\subsection{Relation with previous work}

Our preparation problem has roots in the study of information extraction
without disturbance. In 2002, Koashi and Imoto characterized the operations
that preserve a family of quantum states \cite{KoashiImoto2002}; Kuramochi later
developed this perspective for quantum statistical experiments
\cite{Kuramochi2018}. A complementary approach emerged from Matsumoto's
reverse tests, which characterized the BS relative entropy and maximal
quantum $f$-divergences through preparation from classical distributions
\cite{Matsumoto2010a,Matsumoto2018}. \cite{GourTomamichel2020} subsequently placed
such extensions within a general framework for resource measures. 
We connect these developments by showing that
preserving the preparation memory realizes the maximal extension.
The earlier characterization of entanglement-breaking channels as
measure-and-prepare maps \cite{HorodeckiShorRuskai2003} gives another realization of the
same reverse-preparation optimization.

More recently, \cite{MosonyiBunthVrana2024} used weighted operator means to
interpolate between Umegaki and BS. \cite{CapelCostaRico2026b} and \cite{CapelCostaRico2026a}
studied a geodesic interpolation and its associated classical testing
experiments, which use auxiliary purifications and measurements restricted
to a commutative likelihood algebra.
Here, we obtain an operational interpolation by varying the number of emissions
required from a preparation device.  The resulting hierarchy
$I_n^{\mathbb D}$ quantifies preparation costs. The universal Stein
exponent $r_n$ of the associated testing problem is a separate quantity.
For Umegaki relative entropy, the preparation cost and the testing exponent
agree at $n=1$ and at the BS endpoint as $n\to\infty$; at intermediate
$n \in \mathbb{N}\setminus\{1\}$, we establish the bound $r_n\le I_n^D$, rather than an
identification of the two quantities.

The passage from finite reuse to the BS endpoint draws on the development
of finite quantum de Finetti theorems. \cite{ChristandlKonigMitchisonRenner2007} established quantitative approximations for reduced symmetric
states. The channel version of \cite{Chiribella2011} made
available a common measure-and-prepare approximation valid for all inputs.
We apply it to the average marginal channel, obtaining a measurement
independent of the unknown preparation device. The resulting estimate
connects the finite preparation costs to universal testing and controls
their convergence to BS; its proof is given in
Appendix~\ref{app:uniform-extraction}.

The testing problem continues the study of quantum discrimination beyond
known pairs of states. In \cite{BertaBrandaoHirche2021} it was shown that composite
quantum hypotheses can require regularization. The composite generalized Stein theorem of
 \cite{LamiRegulaTakagi2026} provides the basis for our analysis. A parallel classical development, adversarial hypothesis
testing \cite{BrandaoHarrowLeePeres2020}, supplies the convex-family argument used to reach
the BS endpoint. We retain the distinction between the universal exponent
$r_n$ and the preparation cost $I_n^D$: we prove $r_n\le I_n^D$ at finite
$n$ and show that both approach BS as the number of emissions grows.

\paragraph{Organization.}
Section~\ref{sec:memory} develops the divergence examples and
entanglement-breaking preparation and establishes a BS Stein limit for testing the memory
states of a known reusable source. Section~\ref{sec:extensions} introduces
the finite costs $I_n^{\mathbb D}$ associated to different memory setups, proves data processing, and
relates their limit to indefinite preparation and maximal extensions,
including the quantitative BS convergence. Section~\ref{sec:testing}
studies universal discrimination of emissions from an unknown device,
compares its fixed exponent with $I_n^D$, and proves the BS limit
\eqref{eq:main-intro} with independent resets. Appendix~\ref{app:uniform-extraction}
proves the common extraction estimate used in the robust, the hierarchy and the
testing argument.

\section{Stationary preparations and their information cost}\label{sec:memory}

\subsection{Notation}

All Hilbert spaces are finite-dimensional. We write $\cL(H)$ for the operators
on $H$ and $\D(H)$ for the density operators. We also write
$\mathcal B(H)=\cL(H)$. A channel is a completely positive,
trace-preserving linear map, acting in the Schr\"odinger picture.
We use $\Phi:H\to K$ as shorthand for a channel
$\Phi:\cL(H)\to\cL(K)$. The trace norm
is $\norm{X}_1=\tr\sqrt{X^*X}$. All logarithms are natural. The Umegaki relative
entropy is
\[
 D(\rho \Vert \sigma)=\tr \rho (\log \rho-\log \sigma)
\]
when $\supp \rho \subseteq\supp \sigma$, and is $+\infty$ otherwise
\cite{Umegaki1962}. The same support convention is used for the BS relative entropy,
with inverses taken on $\supp \sigma$ and $0\log0=0$. Data processing for these
divergences is recalled in \cite{Lindblad1975,Matsumoto2010a}. 
The prescribed states $\rho,\sigma$ satisfy the support and
faithfulness assumptions stated in each result; optimized memories
and extensions need not be faithful. We use $\rho_0=\rho$ and $\rho_1=\sigma$ when it is
convenient to treat the hypotheses together.

A general divergence satisfying data processing
is denoted by $\mathbb D$, its classical restriction by
$\mathbb D_{\mathrm{cl}}$, and its maximal extension by
$\widehat{\mathbb D}$, as defined in \eqref{eq:maximal-extension}.
The preparation identities require only support inclusion; faithfulness
of the targets is imposed for the quantitative BS convergence and the
universal output-testing results. Optimized memory states and emitted joint states
need not be faithful, and their supports need not be nested.


\subsection{Divergence examples}\label{subsec:examples}

We can now apply this result to
many different relative entropies. In particular, we can consider  the Umegaki relative entropy,
the R\'enyi entropies, and the hypothesis-testing divergence. The combination of Theorem \ref{thm:preparation} together with the known reverse-tests \cite{Matsumoto2010a,Matsumoto2018,MosonyiBunthVrana2024,GourTomamichel2020} provides the following identifications:

\begin{enumerate}[label=\textup{(\roman*)},leftmargin=2.6em]
\item  For the Umegaki relative entropy, whose classical restriction is the Kullback–Leibler divergence, Matsumoto's reverse-test characterization \cite{Matsumoto2010a} identifies the maximal extension with the Belavkin–Staszewski relative entropy \cite{BelavkinStaszewski1982},
\begin{equation}
 \inf_{\substack{M,\omega_0,\omega_1,\cR:M\to M\otimes S\\
                  \tr_S\cR(\omega_i)=\omega_i\\
                  \tr_M\cR(\omega_i)=\rho_i\ (i=0,1)}}
       D(\omega_0\|\omega_1)
  =D_{\mathrm{BS}}(\rho_0\|\rho_1)\, .
  \label{eq:bs-minimum}
\end{equation}
 The stationary infimum is attained by a classical memory of dimension at
most $\dim S$. 

\item For $\mathbb{D}=D_\alpha$, where $D_{\alpha}$ denotes the Petz Rényi divergence and where
$0<\alpha<1$ or $1<\alpha\le2$, the stationary minimum
is the \emph{geometric R\'enyi divergence} \cite{Matsumoto2018,HiaiMosonyi2017},
\begin{equation}
  \inf_{\substack{M,\omega_0,\omega_1,\cR:M\to M\otimes S\\
                  \tr_S\cR(\omega_i)=\omega_i\\
                  \tr_M\cR(\omega_i)=\rho_i\ (i=0,1)}}
       D_{\alpha}(\omega_0\|\omega_1)
  =\widehat{D}_\alpha(\rho_0\|\rho_1)
  \label{eq:geometric-renyi-minimum}
\end{equation}
The same classical memory as in \textup{(i)} attains the infimum, again
with dimension at most $\dim S$\footnote{For the regularized prepared R\'enyi divergence
when $\alpha>2$, see \cite{Rubboli2026}. That regularized optimization is not
the single-copy reverse-preparation problem considered here.}
. This result can be analogously extended to general quantum $f$-divergences with operator-convex functions $f$ whose classical restriction has the maximal quantum $f$-divergence as its maximal extension  \cite{Matsumoto2018,HiaiMosonyi2017}.

\item The same characterization applies to the trace distance
$T(\rho_0,\rho_1)=\frac12\|\rho_0-\rho_1\|_1$ and the purified
distance $P(\rho_0,\rho_1)=\sqrt{1-\|\sqrt\rho_0\sqrt\rho_1\|_1^2}$.
Their maximal extensions have the explicit expressions
\cite{Matsumoto2018,RegulaLamiBertaTomamichel2026}
\begin{equation}
\begin{aligned}
\widehat T(\rho_0,\rho_1)
&=1-\max_{\substack{0\le X\le\rho_0\\X\le\rho_1}}
  \tr X,\\
\widehat P(\rho_0,\rho_1)
&=\sqrt{1-
 \left(\tr\left[
 \rho_1(\rho_1^{-1/2}\rho_0\rho_1^{-1/2})^{1/2}
 \right]\right)^2}.
\end{aligned}
\end{equation}
By Theorem~\ref{thm:preparation}, these quantities
are precisely the corresponding minimum information
costs of stationary preparations.

\item For $0<\varepsilon<1$ we can consider the  
hypothesis-testing divergence $\mathbb{D}=D_H^\varepsilon$ \cite{WangRenner2012}, given by  
\begin{equation}\label{eq:DHdef}
 D_H^\varepsilon(\rho_0\|\rho_1)
 :=-\log\inf\bigl\{\tr(Q\rho_1):0\le Q\le I,\ 
                      \tr(Q\rho_0)\ge1-\varepsilon\bigr\},
\end{equation}
 The stationary minimum is the \emph{maximal hypothesis-testing
divergence}
\begin{equation}
   \inf_{\substack{M,\omega_0,\omega_1,\cR:M\to M\otimes S\\
                  \tr_S\cR(\omega_i)=\omega_i\\
                  \tr_M\cR(\omega_i)=\rho_i\ (i=0,1)}}
       D_H^\varepsilon(\omega_0\|\omega_1)
 =\widehat D_H^\varepsilon(\rho_0\|\rho_1)
 :=\inf_{\substack{p_0,p_1,\{\tau_z\}\\
               \sum_zp_i(z)\tau_z=\rho_i\ (i=0,1)}}
       D_{H,\mathrm{cl}}^\varepsilon(p_0\|p_1).
 \label{eq:maximal-hypothesis-testing-minimum}
\end{equation}
An explicit semidefinite characterization of the maximal hypothesis-testing divergence was also recently obtained in \cite[Section~IV]{RegulaLamiBertaTomamichel2026}.
\end{enumerate}

In \textup{(iv)}, the preparation theorem gives a variational
identification. It does not require
attainment of the reverse-preparation infimum, and no dimension bound
for an optimal memory is asserted in this case.

\subsection{Entanglement-breaking preparation}

In this section we show that a different setup can yield the same minimum information cost as in Theorem~\ref{thm:preparation}. We will restrict the preparation channel to be entanglement
breaking. A channel $\Phi:M\to S$ is entanglement breaking
if $(\id_A\otimes\Phi)(\xi)$ is separable for every ancillary
system $A$ and every state $\xi$ on $A\otimes M$. Equivalently,
it admits a finite measure-and-prepare representation
\cite[Theorem~2]{HorodeckiShorRuskai2003},
\begin{equation}\label{eq:MP}
 \Phi(X)=\sum_z\tr(E_zX)\tau_z,
 \qquad E_z\ge0,\quad\sum_zE_z=I_M.
\end{equation}

The following result shows that the stationarity constraint on the memory \eqref{eq:stationarity-constraint} can be replaced by the map being entanglement breaking and both restrictions lead to the same minimum information cost.

\begin{proposition}[Entanglement-breaking preparation]\label{prop:EB}
Let $\mathbb{D}$ be a quantum divergence and $\rho_0,\rho_1 \in \mathcal{D}(S)$ with  $
\operatorname{supp}\rho_0\subseteq\operatorname{supp}\rho_1
$. Then,
\begin{equation}\label{eq:EB}
 \inf_{\substack{M,\omega_0,\omega_1,\Phi:M\to S\\
       \Phi\ \mathrm{entanglement\ breaking}\\
       \Phi(\omega_i)=\rho_i\ (i=0,1)}}
 \mathbb{D}(\omega_0\Vert\omega_1)
 =\widehat{\mathbb{D}}(\rho_0\Vert\rho_1).
\end{equation} 
\end{proposition}

\begin{proof}
For a representation \eqref{eq:MP}, put $p_i(z)=\tr(E_z\omega_i)$.
Then $\rho_i=\sum_zp_i(z)\tau_z$. Data processing under the measurement
$X\mapsto\diag(\tr E_zX)_z$ and the definition
\eqref{eq:maximal-extension} give
\[
 \mathbb{D}(\omega_0\Vert\omega_1)
 \ge\mathbb{D}_{\mathrm{cl}}(p_0\Vert p_1)
 \ge\widehat{\mathbb{D}}(\rho_0\Vert\rho_1).
\]
Conversely, every classical reverse preparation $\rho_i=\sum_z p_i(z)\tau_z$ can define the
entanglement-breaking channel
$\Phi(X)=\sum_z\bra zX\ket z\,\tau_z$ with classical input states
$\omega_i=\sum_z p_i(z)\ket z\bra z$ satisfying $\Phi(\omega_i)=\rho_i$. Thus \((M,\omega_0,\omega_1,\Phi)\) is admissible for the optimization and 
$\mathbb{D}(\omega_0\Vert \omega_1)=\mathbb{D}_{\mathrm{cl}}(p_0\Vert p_1)$. Taking the infimum  over admissible quadruples and afterwards over admissible reverse preparations proves
\eqref{eq:EB}. 
\end{proof}

Proposition \ref{prop:EB} shows that entanglement-breaking preparation and stationary preparation have the same minimum information cost. In particular, classical reverse preparations suffice to approximate this infimum arbitrarily closely. However, this does not mean that correlations between the retained memory and the emitted system can be excluded altogether. Indeed, suppose we impose the stronger condition
\[
\mathcal R(\omega_i)=\omega_i\otimes\rho_i,\qquad i=0,1.
\]
Monotonicity of fidelity under channels and its multiplicativity give
\[
F(\omega_0,\omega_1)
\le F(\omega_0,\omega_1)F(\rho_0,\rho_1).
\]
For distinct target states $\rho_0\ne\rho_1$, this forces $F(\omega_0,\omega_1)=0$, so the memory states have orthogonal supports and $D(\omega_0\Vert\omega_1)=+\infty$. Consequently, while entanglement is unnecessary to attain the optimal preparation cost, complete decoupling is incompatible with finite relative-entropy cost. Correlations are therefore essential, but classical correlations suffice.

\subsection{Hypothesis testing of stationary memories}\label{subsec:source-discrimination}

The previous subsections show that the minimum divergence of a stationary
preparation is the maximal extension of the corresponding classical
divergence. For the Umegaki relative entropy, this minimum is the
Belavkin--Staszewski relative entropy. We now give this value a
hypothesis-testing interpretation in terms of the memory states themselves.

Fix a stationary preparation $(M,\omega_0,\omega_1,\cR)$ of
$\rho_0,\rho_1$. Instead of observing the emitted systems, suppose the
tester receives $n$ independently initialized copies of the memory, in
one of the two states $\omega_0^{\otimes n}$ or $\omega_1^{\otimes n}$.
The preparation device is known, so the discrimination test may depend
on the particular memory states. Since different stationary preparations
can produce the same prescribed outputs, we minimize the optimal
testing divergence over all compatible sources. Let $D_H^{\varepsilon}(\rho_0\Vert \rho_1)$ denote the hypothesis-testing relative entropy. 
 We define 
 \begin{equation}\label{eq:sourceDH}
  D_{H}^{\varepsilon,n}(\rho_0\Vert\rho_1)
  :=
\inf_{\substack{p_0,p_1,\{\tau_z\}\\
\rho_i=\sum_zp_i(z)\tau_z}}
D_H^\varepsilon
(p_0^{\otimes n}\Vert p_1^{\otimes n})
=
  \inf_{\substack{M,\omega_0,\omega_1,\cR:M\to M\otimes S\\
                  \tr_S\cR(\omega_i)=\omega_i\\
                  \tr_M\cR(\omega_i)=\rho_i\ (i=0,1)}}
  D_H^{\varepsilon}(\omega_0^{\otimes n}\Vert\omega_1^{\otimes n}).
 \end{equation}

The equality in \eqref{eq:sourceDH} follows by applying
Theorem~\ref{thm:preparation} to the divergence
$\mathbb D_n(a\Vert b)=D_H^\varepsilon(a^{\otimes n}\Vert b^{\otimes n})$.
It satisfies data processing because a common channel on $a,b$ acts
as its tensor power on the corresponding $n$-copy states.

The $n$ copies in \eqref{eq:sourceDH} are independent initializations of
the memory, not $n$ successive emissions from one initialization. For $n=1$, this definition coincides with the maximal hypothesis-testing divergence \eqref{eq:maximal-hypothesis-testing-minimum}. For
each fixed preparation, the quantum Stein lemma determines the
asymptotic exponent \cite{HiaiPetz1991,OgawaNagaoka2000}
\[
\lim_{n\to\infty}\frac1n
D_H^\varepsilon
\left(
\omega_0^{\otimes n}\Vert\omega_1^{\otimes n}
\right)
=
D(\omega_0\Vert\omega_1)\, .
\]
Since we minimize over
all compatible preparations, however, the limit cannot simply be
interchanged with the infimum. Showing that this permutation of the  limit and the infimum is possible is the goal of the next result.

\begin{theorem}[Stein limit for stationary memories]\label{thm:source-Stein}
Let $\rho_0,\rho_1\in\D(S)$ satisfy
$\supp\rho_0\subseteq\supp\rho_1$. Then, for every fixed
$0<\varepsilon<1$,
\begin{equation}\label{eq:source-Stein}
 \lim_{n\to\infty}\frac1n
 D_{H}^{\varepsilon,n}(\rho_0\Vert\rho_1)
 =\BS(\rho_0\Vert\rho_1).
\end{equation}
\end{theorem}
\begin{proof}
    The proof can be found in Theorem \ref{AP-thm:source-Stein}.
\end{proof}

A closely related asymptotic result was established by Regula, Lami, Berta and Tomamichel \cite[Proposition~2]{RegulaLamiBertaTomamichel2026}, who proved that, for every fixed $0<\delta<1$,
\[
 \lim_{n\to\infty}\frac1n
 \widehat D_{\max}^{\delta}
 (\rho_0^{\otimes n}\Vert\rho_1^{\otimes n})
 =\BS(\rho_0\Vert\rho_1).
\]
Here $\widehat D_{\max}^{\delta}$ denotes the maximal extension of the classical smoothed max-relative entropy,
\[
 D_{\max}^{\delta,\mathrm{cl}}(p\Vert q)
 =\log\inf\left\{\lambda>0:
 \sum_z(p_z-\lambda q_z)_+\le\delta\right\}.
\]
More precisely, their optimization is over all classical reverse preparations of the tensor-product pair $(\rho_0^{\otimes n},\rho_1^{\otimes n})$, allowing the classical distributions and preparation channel to depend jointly on $n$. In contrast, our quantity $D_H^{\varepsilon,n}$ 
first optimizes over single-copy reverse preparations and then tests $n$ independent copies of the corresponding records. Although these optimizations differ for finite $n$, both have the same asymptotic rate, given by the Belavkin--Staszewski relative entropy.

In the testing problem considered here, the preparation device is known to
the tester, who has direct access to independently initialized copies of its
memory. The test may therefore be chosen for the particular pair of memory
states, and only afterwards do we minimize over all compatible stationary
preparations. In this sense, the uncertainty about the preparation enters
through the optimization of the discrimination rate, not through the test
itself.

In Section~\ref{sec:testing}, we will introduce a more general testing
scenario in which the device is unknown and the tester observes only its
emitted systems. We will no longer require the memory to remain stationary:
the device need only produce the prescribed outputs for a finite number of
uses, and these outputs may be correlated. The tester must then choose a
single test that works for every compatible device. We will show that, as
the number of emissions per initialization increases, the resulting universal
Stein exponent also converges to $\BS(\rho_0\Vert\rho_1)$. Thus the same
divergence arises both from testing a known reusable memory and from testing
the outputs of an unknown preparation device.

\section{Repeated quantum preparation and its information cost}\label{sec:extensions}

\subsection{Sequential preparation}
\label{subsec:sequential-preparation}

In Theorem~\ref{thm:preparation}, we showed that maximal quantum divergences arise as the minimum information cost of a preparation device that preserves its memory state after emitting the prescribed output. The fixed-point condition \eqref{eq:stationarity-constraint} ensures that the same device can be used indefinitely, but it imposes a  constraint on the retained memory rather than on the emitted systems. We now show that this condition can be relaxed without changing the optimal information cost: it suffices to require that the device produce the prescribed output at every step of an infinite sequence of uses, while allowing the memory to evolve.

To formulate this relaxation, we consider repeated applications of the same preparation channel $\mathcal R:M\to M\otimes S$ from Figure \ref{fig:memory}, feeding the retained memory into the next application and leaving previously emitted systems untouched. Writing
\begin{equation}\label{eq:horizon-marginal-channels}
\mathcal T=\operatorname{Tr}_S\circ\mathcal R,
\qquad
\mathcal O=\operatorname{Tr}_M\circ\mathcal R,
\end{equation}
the marginal of the $(k+1)$-st emitted system is $\mathcal O(\mathcal T^k(\omega_i))$. The following proposition establishes that requiring these marginals to equal $\rho_i$ for every $k\ge0$ leads to the same minimum as stationary preparation, even though the memory itself need not remain invariant.

\begin{proposition}[Infinite memory chain]
\label{prop:output-stationarity}\label{cor:output-stationarity}
Let $\rho_0,\rho_1\in\D(S)$ satisfy
$\supp\rho_0\subseteq\supp\rho_1$, and let $\mathbb D$  
be a quantum divergence. Then,
\begin{equation}\label{eq:output-stationarity-minimum}
 \inf_{\substack{M,\omega_0,\omega_1,\cR:M\to M\otimes S\\
                  \mathcal O(\mathcal T^k(\omega_i))=\rho_i\\
                  i=0,1,\ k\ge0}}
       \mathbb D(\omega_0\Vert\omega_1)\\
=\widehat{\mathbb D}(\rho_0\Vert\rho_1)\, .
\end{equation}
\end{proposition}

\begin{proof}
The proof can be found in Proposition~\ref{AP-prop:output-stationarity}.
\end{proof}

\begin{figure}[htbp]
\centering
\begin{tikzpicture}[>=Latex,font=\small,x=1cm,y=1cm,
                    every path/.style={draw=black},every node/.style={text=black}]
\node[anchor=east] at (0,.2) {$\omega_i$};
\foreach \x in {1.15,4.15,8.35}{
 \draw[rounded corners=1pt] (\x-.45,-.25) rectangle (\x+.45,.65);
 \node at (\x,.2) {$\cR$};
 \draw[->] (\x,-.25)--(\x,-1.1);
}
\draw[->] (0,.2)--(.7,.2);
\draw[->] (1.6,.2)--(3.7,.2);
\node[above] at (2.65,.3) {$\mathcal T(\omega_i)$};
\draw[->] (4.6,.2)--(5.65,.2);
\node at (6.15,.2) {$\cdots$};
\draw[->] (6.7,.2)--(7.9,.2);
\node[above] at (6.7,.65) {$\mathcal T^{n-1}(\omega_i)$};
\draw[->] (8.8,.2)--(10.45,.2);
\node[above] at (9.65,.3) {$\mathcal T^n(\omega_i)$};
\node[below,align=center] at (1.15,-1.1) {$S_1$\\$\rho_i$};
\node[below,align=center] at (4.15,-1.1) {$S_2$\\$\rho_i$};
\node[below,align=center] at (8.35,-1.1) {$S_n$\\$\rho_i$};
\end{tikzpicture}
\caption{Sequential composition of the preparation block in
Figure~\ref{fig:memory}. The retained-memory output of one use is the input
of the next, with the same channel $\cR$ throughout. Only the first $n$
emitted marginals are prescribed; the memory need not be stationary.}
\label{fig:finite-horizon}\label{fig:extensions}
\end{figure}
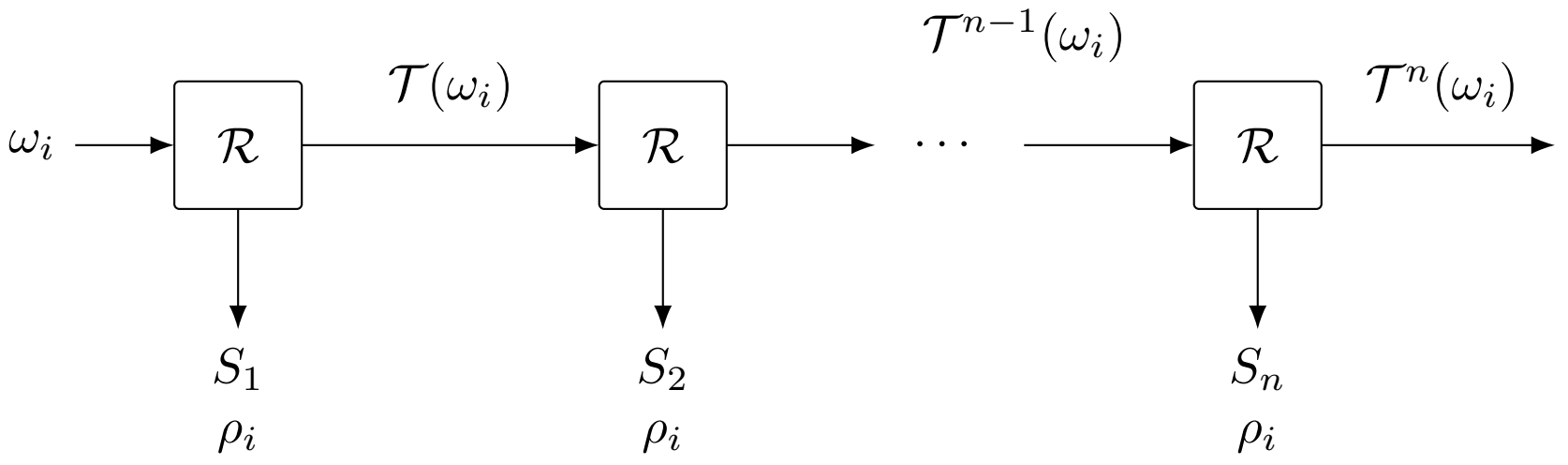

The previous result raises a natural question: how does the information cost change if the device is required to produce the prescribed states for only a finite number of uses? To address this question, we consider the same preparation channel applied successively $n$ times, as illustrated in Figure~\ref{fig:finite-horizon}. We require the first $n$ emitted systems to have the prescribed marginals, without imposing any condition on the evolution of the retained memory. This leads to the following $n$-use preparation cost, which reduces to unrestricted preparation for $n=1$. Its convergence to the indefinite-use preparation cost will be established below.

\begin{definition}\label{def:I_n}
Let $n\ge1$ and let $\mathbb D$ be a quantum divergence. For $\rho_0,\rho_1 \in \mathcal{D}(S)$, define the \emph{$n$-use preparation cost}
\begin{equation}\label{eq:finite-horizon-cost}
 I_n^{\mathbb D}(\rho_0\Vert\rho_1)
 :=
 \inf_{\substack{M,\omega_0,\omega_1,\cR:M\to M\otimes S\\
       \mathcal O(\mathcal T^k(\omega_i))=\rho_i\\
       i=0,1,\quad k=0,\ldots,n-1}}
       \mathbb D(\omega_0\Vert\omega_1).
\end{equation}
The infimum is over finite-dimensional memories, arbitrary memory states,
and channels common to both hypotheses. The memory and device may depend
on $n$; no memory-dimension bound uniform in $n$ is imposed. We write
$I_\infty^{\mathbb D}$ for the \emph{indefinite-use preparation cost}, which corresponds to the same infimum with the output conditions
imposed for every $k\ge0$.
\end{definition}

For $n=1$ and $n=\infty$, Theorem~\ref{thm:preparation} and Proposition~\ref{prop:output-stationarity} show that the single-use preparation cost and the indefinite-use preparation cost correspond to the quantum divergence and its maximal extension respectively, i.e.
\begin{equation}\label{eq:horizon-endpoints}
     I_1^{\mathbb D}(\rho_0\Vert\rho_1)= \mathbb D(\rho_0\Vert\rho_1) \, , \qquad I_\infty^{\mathbb D}(\rho_0\Vert\rho_1)
     =\widehat{\mathbb D}(\rho_0\Vert\rho_1)\, . 
\end{equation}
The next result shows that actually $I_n^{\mathbb D}$ is a quantum divergence and has the same classical restriction as $\mathbb{D}$ for every $n \in \mathbb{N}\cup \{+\infty\}$.

\begin{theorem}[Data processing and classical restriction]
\label{thm:horizon-divergences}
Let $\mathbb D$ be a quantum divergence with classical
restriction $\mathbb{D}_{\mathrm{cl}}$, and let
$\rho_0,\rho_1\in\D(S)$ satisfy
$\supp\rho_0\subseteq\supp\rho_1$. For every
$n\in\mathbb N\cup\{+\infty\}$ and every channel $\Phi:S\to S'$, one has
\begin{equation}\label{eq:horizon-dpi}
 I_n^{\mathbb D}\bigl(\Phi(\rho_0)\Vert\Phi(\rho_1)\bigr)
 \le I_n^{\mathbb D}(\rho_0\Vert\rho_1).
\end{equation}
Moreover, for probability distributions $p,q$ with
$\supp p\subseteq\supp q$,
\begin{equation}\label{eq:horizon-classical-restriction}
 I_n^{\mathbb D}(\diag p\Vert\diag q)
   =\mathbb D_{\mathrm{cl}}(p\Vert q).
\end{equation}
\end{theorem}

\begin{proof}
The proof can be found in Theorem~\ref{AP-thm:horizon-divergences}.
\end{proof}

\begin{remark}
Notice that the proof of data-processing inequality  of $I_n^{\mathbb{D}}$ does not use data processing of the quantum divergence $\mathbb{D}$, it follows
from closure of the preparation model under output postprocessing. However, the identifications
$I_1^{\mathbb D}=\mathbb D$ or $I_{\infty}^{\mathbb D}=\widehat{\mathbb D}$ do rely on
this assumption. Without it,
$I_n^{\mathbb D}$ still satisfies data
processing, but the single-use  or the indefinitely-use cost need not
coincide with the original functionals
$\mathbb D$ and $\widehat{\mathbb D}$.
\end{remark}

\subsection{Parallel preparation}
\label{subsec:parallel-preparation}

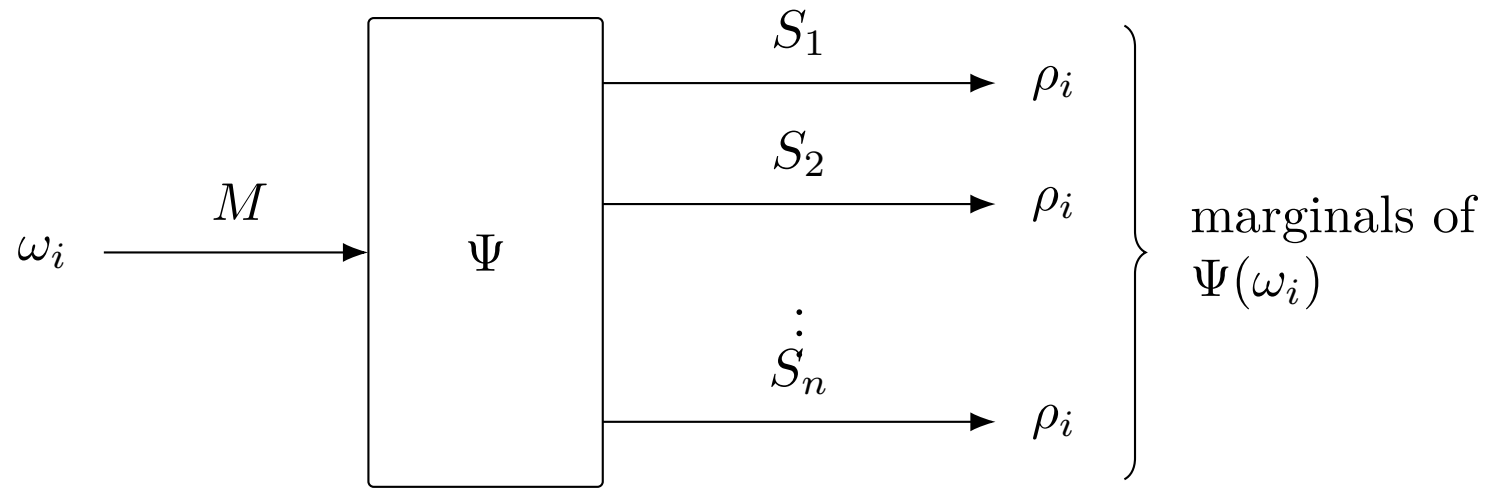
\begin{figure}[htbp]
\centering
\begin{tikzpicture}[>=Latex,font=\small,x=1cm,y=1cm,
                    every path/.style={draw=black},every node/.style={text=black}]
\draw[rounded corners=1pt] (2.5,-1.55) rectangle (4.05,1.55);
\node at (3.275,0) {$\Psi$};
\node[anchor=east] at (.65,0) {$\omega_i$};
\draw[->] (.75,0)--(2.5,0);
\node[above] at (1.65,.08) {$M$};
\draw[->] (4.05,1.12)--(6.65,1.12);
\draw[->] (4.05,.32)--(6.65,.32);
\draw[->] (4.05,-1.12)--(6.65,-1.12);
\node[above] at (5.35,1.17) {$S_1$};
\node[above] at (5.35,.37) {$S_2$};
\node[above] at (5.35,-1.07) {$S_n$};
\node[anchor=west] at (6.75,1.12) {$\rho_i$};
\node[anchor=west] at (6.75,.32) {$\rho_i$};
\node[anchor=west] at (6.75,-1.12) {$\rho_i$};
\node at (5.35,-.43) {$\vdots$};
\draw[decorate,decoration={brace,amplitude=4pt}]
 (7.5,1.5)--(7.5,-1.5);
\node[anchor=west,align=left] at (7.8,0)
 {marginals of\\$\Psi(\omega_i)$};
\end{tikzpicture}
\caption{Parallel preparation from an initial memory
state $\omega_i$. The channel
$\Psi:M\to S_1\otimes\cdots\otimes S_n$
outputs only the emitted systems; no retained-memory
output is included in this description. For a
sequential device, this channel is obtained as
$\Psi_n=\tr_M\circ\cR^{[n]}$ by discarding the final
memory. The labels $\rho_i$ specify the one-site
marginals of $\Psi(\omega_i)$, not a product state.
Proposition~\ref{prop:parallel-preparation} shows
that the parallel and sequential models have the
same minimum input cost $I_n^{\mathbb D}$.}
\label{fig:parallel-preparation}
\end{figure}

The sequential preparation introduced in Section~\ref{subsec:sequential-preparation} consists
of repeated applications of the same channel $\cR$, with the
retained memory connecting consecutive uses, as illustrated
in Figure~\ref{fig:finite-horizon}. If we discard the final
memory and collect the $n$ emitted systems, the entire
sequence can instead be represented by a single channel
from the initial memory to $S^{\otimes n}$. This is the
description illustrated in Figure~\ref{fig:parallel-preparation},
where the intermediate memory connections are no longer
displayed. Importantly, the emitted systems need not be
independent and may contain arbitrary correlations.

More precisely, let
$\cR^{[n]}:M\to M\otimes S^{\otimes n}$ denote the channel
obtained by $n$ successive applications of $\cR$, leaving
previously emitted systems untouched. The corresponding
output channel is
\begin{equation}\label{eq:sequential-output-channel}
 \Psi_n:=\tr_M\circ\cR^{[n]}.
\end{equation}
For every initial memory state $X$, the retained memory
after $n$ uses and the individual output marginals satisfy
\begin{equation}\label{eq:sequential-output-marginals}
 \begin{aligned}
 \tr_{S^{\otimes n}}\cR^{[n]}(X)
   &=\mathcal T^n(X),\\
 [\Psi_n(X)]_{S_j}
   &=\mathcal O(\mathcal T^{j-1}(X)),
   \qquad 1\le j\le n.
 \end{aligned}
\end{equation}
In particular, for every admissible $n$-use preparation,
each marginal of $\Psi_n(\omega_i)$ equals $\rho_i$.

This observation motivates a more general description in
which the entire output block is prepared by an arbitrary
common channel
$\Psi:M\to S_1\otimes\cdots\otimes S_n$, with $S_j\simeq S$,
satisfying
\begin{equation}\label{eq:parallel-marginal-condition}
 [\Psi(\omega_i)]_{S_j}=\rho_i,
 \qquad i=0,1,\quad 1\le j\le n.
\end{equation}
Unlike the preceding construction, $\Psi$ is not assumed
to arise from successive applications of a memory channel.

The marginal conditions in \eqref{eq:parallel-marginal-condition}
resemble those of quantum broadcasting, but there is an important
distinction: the input records $\omega_i$ are not required to
coincide with the target states $\rho_i$. Ordinary broadcasting
is recovered only by imposing $M=S$ and $\omega_i=\rho_i$,
which, for $n\ge2$, is possible precisely when the target states
commute \cite{BarnumCavesFuchsJozsaSchumacher1996}. Here, instead, the input
records and the preparation channel are optimized, while
arbitrary correlations among the emitted systems are allowed.

The parallel formulation can also be described directly in
terms of its joint output states. For $a\in\D(S)$, define
\begin{equation}\label{eq:extensions}
 \cE_n(a)=\{X\in\D(S^{\otimes n}):
                 \tr_{S_{\ne j}}X=a,\ 1\le j\le n\}
\end{equation}
as the set of compatible $n$-system extensions of $a$.
Every admissible parallel preparation produces a pair of
states in $\cE_n(\rho_0)\times\cE_n(\rho_1)$, with no
restriction on their correlations. Conversely, every such
pair can be realized by a common block-preparation channel.
The following proposition shows that these output pairs
can also be generated by successive applications of a
single preparation channel, so that the sequential, parallel,
and extension formulations yield the same minimum
information cost.
\begin{proposition}[Parallel preparation and extensions]
\label{prop:parallel-preparation}
Let $n\ge1$, let $\rho_0,\rho_1\in\D(S)$ with $\operatorname{supp}\rho_0\subseteq\operatorname{supp}\rho_1$, and let $\mathbb D$ satisfy
data processing. Then
\begin{equation}\label{eq:finite-use}
 I_n^{\mathbb D}(\rho_0\Vert\rho_1)
  =\inf_{\substack{M,\omega_0,\omega_1,\Psi:M\to S^{\otimes n}\\
          [\Psi(\omega_i)]_{S_j}=\rho_i\\
          i=0,1,\ 1\le j\le n}}
        \mathbb D(\omega_0\Vert\omega_1),
\end{equation}
and
\begin{equation}\label{eq:horizon-extension-equivalence}
 I_n^{\mathbb D}(\rho_0\Vert\rho_1)
  =\inf_{\substack{R_n\in\cE_n(\rho_0)\\T_n\in\cE_n(\rho_1)}}
       \mathbb D(R_n\Vert T_n).
\end{equation}
Moreover, if \(\mathbb D\) is jointly lower semicontinuous and its classical restriction is finite for all pairs of probability distributions with support inclusion, then, under \(\operatorname{supp}\rho_0\subseteq\operatorname{supp}\rho_1\), all three infima are finite and attained. A memory of dimension \((\dim S)^n\) suffices.
\end{proposition}

\begin{proof}
The proof can be found in Proposition~\ref{AP-prop:parallel-preparation}.
\end{proof}

\begin{remark}
The optimization over compatible extensions in
\eqref{eq:horizon-extension-equivalence} is closely
related to the work of Horodecki et al.
\cite{Horodecki2006}, who studied
the increase of ensemble information measures,
in particular the Holevo information, under
correlated extensions with prescribed marginals.
Here, we interpret this optimization as the minimum
information cost of sequential and parallel quantum preparation.
\end{remark}

The next result shows that the $n$-use preparation costs provide an interpolation between a quantum divergence and its maximal extension. Related interpolations have been studied through geometric and barycentric constructions by Mosonyi, Bunth and Vrana \cite{MosonyiBunthVrana2024}, and through geodesic quantum $f$-divergences by some authors of this work \cite{CapelCostaRico2026a}. In contrast to the previous constructions, the present hierarchy has an operational interpretation: its parameter is the number of times a preparation device must reproduce the prescribed states. The following proposition establishes monotonicity and convergence to the maximal extension.

\begin{theorem}[Interpolation and convergence of $n$-use preparation costs]
\label{prop:hierarchy}\label{thm:hierarchy}
Let $\mathbb D$ be a quantum divergence and let
$\rho_0,\rho_1\in\D(S)$ with $\supp\rho_0\subseteq\supp\rho_1$.
For every $n\ge1$,
\begin{equation}\label{eq:horizon-general-hierarchy}
 \mathbb D(\rho_0\Vert\rho_1)=I_1^{\mathbb D}(\rho_0\Vert\rho_1) \le
 I_n^{\mathbb D}(\rho_0\Vert\rho_1)
   \le I_{n+1}^{\mathbb D}(\rho_0\Vert\rho_1)
    \le I_\infty^{\mathbb D}(\rho_0\Vert\rho_1)
    =\widehat{\mathbb D}(\rho_0\Vert\rho_1).
\end{equation}
If, in addition, $\widehat{\mathbb D}$ is jointly lower
semicontinuous at $(\rho_0,\rho_1)$ for pairs of states supported on
$\supp\rho_1$, then
\begin{equation}\label{eq:hierarchy}
 \lim_{n\to\infty}I_n^{\mathbb D}(\rho_0\Vert\rho_1)
   =I_\infty^{\mathbb D}(\rho_0\Vert\rho_1)
   =\widehat{\mathbb D}(\rho_0\Vert\rho_1).
\end{equation}
\end{theorem}

\begin{proof}
The proof can be found in Theorem~\ref{AP-thm:hierarchy}.
\end{proof}

\subsection{Strict interpolation and recoverability}
\label{subsec:strict-interpolation}

The preceding theorem shows that the preparation costs increase towards the
maximal extension, but does not rule out equality at a finite number of uses.
We now show that, for divergences whose equality in data processing implies
exact recovery, such an equality can occur only for commuting target states.

\begin{proposition}[Strict interpolation from recoverability]
\label{prop:strict-hierarchy-general}
Let $\mathbb D$ be a quantum divergence, and let
$\rho_0,\rho_1\in\D(S)$ satisfy
$\supp\rho_0\subseteq\supp\rho_1$ and
$\widehat{\mathbb D}(\rho_0\Vert\rho_1)<\infty$.
Assume that the infimum defining $I_k^{\mathbb D}(\rho_0\Vert\rho_1)$
is attained for every finite $k$, and that $\mathbb D$ has the
following recovery property: whenever a channel $\Lambda$ and states
$a_0,a_1$ satisfy
\[
 \mathbb D(\Lambda(a_0)\Vert\Lambda(a_1))
 =\mathbb D(a_0\Vert a_1)<\infty,
\]
there exists a recovery channel $\mathcal Q$ with
$\mathcal Q\circ\Lambda(a_0)=a_0$ and $\mathcal Q\circ\Lambda(a_1)=a_1$.
Then, for any distinct $n,m\in\mathbb N\cup\{+\infty\}$,
\begin{equation}\label{eq:strict-general-hierarchy}
 I_n^{\mathbb D}(\rho_0\Vert\rho_1)
 =I_m^{\mathbb D}(\rho_0\Vert\rho_1)
 \quad\Longleftrightarrow\quad [\rho_0,\rho_1]=0.
\end{equation}
\end{proposition}

\begin{proof}
The proof can be found in Proposition~\ref{AP-prop:strict-hierarchy-general}.
\end{proof}

For Umegaki relative entropy, the attainment assumption
follows from Proposition~\ref{prop:parallel-preparation}
under support inclusion, while equality in data processing
implies recovery \cite{Petz1986,Petz1988}. The same conclusion
extends to standard quantum $f$-divergences for which equality
in data processing characterizes recoverability  \cite[Theorem~3.18]{HiaiMosonyi2017}.  In contrast, equality in data processing for
maximal quantum $f$-divergences does not generally imply
recovery, as demonstrated in
\cite[Section~4.2]{HiaiMosonyi2017}. We extend now this statement to lower semicontinuous maximal divergences.

\begin{proposition}[Failure of recoverability for maximal divergences]
\label{cor:maximal-recoverability}
Let $\widehat{\mathbb D}$ be a lower semicontinuous
maximal divergence. For every pair of noncommuting
states $\rho_0,\rho_1 \in \mathcal{D}(S)$ with $\supp\rho_0\subseteq\supp\rho_1$ and 
$\widehat{\mathbb D}(\rho_0\Vert\rho_1)<\infty$, 
there exist compatible extensions $
R_2\in\mathcal E_2(\rho_0)$, $T_2\in\mathcal E_2(\rho_1)$
such that, for $\Phi=\tr_{S_2}$,
\[
\widehat{\mathbb D}(\Phi(R_2)\Vert\Phi(T_2))
=\widehat{\mathbb D}(R_2\Vert T_2)
=\widehat{\mathbb D}(\rho_0\Vert\rho_1)<\infty,
\]
but no quantum channel $\mathcal Q$ satisfies
\[
\mathcal Q\circ\Phi(R_2)=R_2,
\qquad
\mathcal Q\circ\Phi(T_2)=T_2.
\]
\end{proposition}

\begin{proof}
The proof can be found in Proposition~\ref{AP-cor:maximal-recoverability}.
\end{proof}

\section{Discrimination of \texorpdfstring{$n$}{n}-use preparations}\label{sec:testing}

\subsection{Testing an unknown preparation}

The preceding sections considered the information required in an initial
memory to prepare prescribed output states. We now ask what a receiver can
learn from the emitted systems when neither that memory nor the preparation
device is accessible. This differs from the memory discrimination problem
of Section~\ref{subsec:source-discrimination}, where the record states are
known to the tester. Here the test must be chosen using only the prescribed
output marginals, and it must work for every compatible device. The
sequential, parallel, and extension descriptions of Section~\ref{sec:extensions}
provide equivalent ways to specify this uncertainty.

Fix faithful states $\rho_0$ and $\rho_1$, and a finite use
$n\ge1$. As in Definition~\ref{def:I_n}, let $\mathfrak P_n(\rho_0,\rho_1)$
consist of all finite-dimensional preparations
$\mathsf P=(M,\omega_0,\omega_1,\cR)$ with a common channel
$\cR:M\to M\otimes S$ satisfying
\begin{equation}\label{eq:testing-device-class}
 \mathcal O(\mathcal T^k(\omega_i))=\rho_i,
 \qquad i=0,1,\quad k=0,\ldots,n-1,
\end{equation}
where $\mathcal T=\tr_S\circ\cR$ and $\mathcal O=\tr_M\circ\cR$.
The device is used $n$ times in succession, with its retained memory
fed into the next use; no stationarity condition is imposed. Writing
$\cR^{[n]}$ for this sequential composition, its block-output channel
and the two possible emitted states are
\begin{equation}\label{eq:testing-device-output}
 \Psi_n^{\mathsf P}=\tr_M\circ\cR^{[n]},
 \qquad
 \Omega_{i,n}^{\mathsf P}=\Psi_n^{\mathsf P}(\omega_i),
 \quad i=0,1.
\end{equation}
Each one-site marginal of $\Omega_{i,n}^{\mathsf P}$ equals $\rho_i$,
while the systems emitted at different uses may be correlated.

In the parallel description, $\Psi_n^{\mathsf P}$ is simply a channel
preparing an entire block $S^{\otimes n}$ from one initial record.
The extension description retains only the resulting joint states, as it is shown in the next result.

\begin{lemma}[Equivalence of sequential outputs and compatible extensions]
\label{lem:device-output-pairs}
Let $\rho_0,\rho_1\in\D(S)$ and $n\geq1$.
The pairs of joint output states generated by admissible
$n$-use sequential preparations coincide with the pairs
of compatible extensions. More precisely,
\begin{equation}\label{eq:device-output-pairs}
\bigl\{(\Omega_{0,n}^{\mathsf P},\Omega_{1,n}^{\mathsf P}):
\mathsf P\in\mathfrak P_n(\rho_0,\rho_1)\bigr\}
=\cE_n(\rho_0)\times\cE_n(\rho_1),
\end{equation}
where $\cE_n$ is defined in \eqref{eq:extensions}.
\end{lemma}

\begin{proof}
Let $\mathsf P=(M,\omega_0,\omega_1,\cR)$ be an
admissible sequential preparation, and write
$\mathcal T=\tr_S\circ\cR$ and
$\mathcal O=\tr_M\circ\cR$.
By the marginal constraints, the $k$-th marginal
of its joint output state satisfies
\[
\tr_{S_{\neq k}}\Omega_{i,n}^{\mathsf P}
=\mathcal O(\mathcal T^{k-1}(\omega_i))
=\rho_i,\qquad k=1,\ldots,n.
\]
Thus $\Omega_{i,n}^{\mathsf P}\in\cE_n(\rho_i)$,
which proves the first inclusion.

Conversely, take arbitrary states
$X_i\in\cE_n(\rho_i)$, $i=0,1$.
Following the shift-and-emit construction in the
proof of Proposition~\ref{prop:parallel-preparation},
choose $M=S^{\otimes n}$ and $\omega_i=X_i$.
The common preparation channel $\cR$ emits the
first memory register, shifts the remaining
registers forward, and appends a fixed state
$\chi\in\D(S)$ to the memory. Since this operation
only permutes the original registers and introduces
an independent ancillary state, the joint state of
the first $n$ emitted systems is exactly $X_i$.
In particular,
\[
\Omega_{i,n}^{\mathsf P}=X_i,
\qquad
\mathcal O(\mathcal T^{k-1}(\omega_i))
=\tr_{S_{\neq k}}X_i=\rho_i
\]
for every $k=1,\ldots,n$.
Hence the same channel $\cR$ realizes both
compatible extensions as an admissible sequential
preparation, proving the reverse inclusion.
\end{proof}

A single \emph{round} consists of initializing the device in $\omega_i$,
collecting its $n$ emissions, and discarding the final memory. Before the
next round, the same device is initialized independently in a fresh copy
of $\omega_i$. After $m$ rounds the hypotheses are therefore
\begin{equation}\label{eq:hypotheses}
 \begin{aligned}
 H_0 &: (\Omega_{0,n}^{\mathsf P})^{\otimes m},\\
 H_1 &: (\Omega_{1,n}^{\mathsf P})^{\otimes m},
 \end{aligned}
 \qquad \mathsf P\in\mathfrak P_n(\rho_0,\rho_1).
\end{equation}
The device is fixed across rounds and is common to both hypotheses;
only its initial memory state depends on $i$. It is unknown to the
receiver, however, and the admissible device may be different for
different choices of $n$ and $m$. Independent initialization gives the
tensor-product structure across rounds, not across the $n$ emissions
within a round.

A binary test is an effect $0\le Q\le I$ on $S^{\otimes nm}$, with $Q$
corresponding to acceptance of $H_0$. Its worst-case type-I and type-II
errors are
\begin{align}
 \alpha_{n,m}(Q)
 &=\sup_{\mathsf P\in\mathfrak P_n(\rho_0,\rho_1)}
   \tr\!\left[(I-Q)(\Omega_{0,n}^{\mathsf P})^{\otimes m}\right],
   \label{eq:alpha}\\
 \beta_{n,m}(Q)
 &=\sup_{\mathsf P\in\mathfrak P_n(\rho_0,\rho_1)}
   \tr\!\left[Q(\Omega_{1,n}^{\mathsf P})^{\otimes m}\right].
   \label{eq:beta}
\end{align}
For $0<\varepsilon<1$, define the optimal universal type-II error by
\begin{equation}\label{eq:beta-opt}
 \beta_{n,m}^{\varepsilon}
 :=\inf_{\substack{0\le Q\le I\\\alpha_{n,m}(Q)\le\varepsilon}}
     \beta_{n,m}(Q).
\end{equation}
The same test must satisfy the error constraint for every device; it
may act collectively on all $nm$ output systems but cannot depend on
which preparation produced them. By \eqref{eq:device-output-pairs}, the
suprema may equivalently be taken over the corresponding sets of
compatible extensions.

For $n=1$ these sets contain only $\rho_0$ and
$\rho_1$, and the problem reduces to ordinary quantum hypothesis
testing. For larger $n$, we investigate the Stein exponent
$-m^{-1}\log\beta_{n,m}^{\varepsilon}$ as the number of independently
initialized rounds grows, and then compare it with the preparation
costs $I_n^D$ as $n$ increases. In this section $D$ will denote the Umegaki relative entropy.

\subsection{Universal testing and preparation cost}
\label{subsec:testing-cost}

We first keep the number $n$ of uses in each round fixed and let the
number $m$ of independent initializations increase. Each preparation
then produces a fixed pair of joint output states, but the receiver
knows only their one-site marginals. To relate this testing problem to
Section~\ref{sec:extensions}, we distinguish the information present in
the emitted pair from the rate attainable without knowing that pair.

Proposition~\ref{prop:parallel-preparation} and
Lemma~\ref{lem:device-output-pairs} give, for every divergence
$\mathbb D$ satisfying data processing,
\begin{equation}\label{eq:output-cost-identity}
 \begin{aligned}
 I_n^{\mathbb D}(\rho_0\Vert\rho_1)
 &=\inf_{\mathsf P\in\mathfrak P_n(\rho_0,\rho_1)}
       \mathbb D(\Omega_{0,n}^{\mathsf P}\Vert
                         \Omega_{1,n}^{\mathsf P})\\
 &=\inf_{\substack{R\in\cE_n(\rho_0)\\T\in\cE_n(\rho_1)}}
       \mathbb D(R\Vert T).
 \end{aligned}
\end{equation}
Thus the minimum information cost of an initial record is also the
minimum divergence of an admissible emitted pair. No information is
lost in passing to the output description at the level of this
optimization: the shift-and-emit construction can transfer any pair of
compatible extensions from the memory to the outputs unchanged. For
Umegaki relative entropy, the infimum is finite and attained by
Proposition~\ref{prop:parallel-preparation}.

If the device were known, its emitted pair would be known as well, and
the quantum Stein lemma would give
\begin{equation}\label{eq:known-device-Stein}
 \lim_{m\to\infty}\frac1m
 D_H^\varepsilon\!\left(
   (\Omega_{0,n}^{\mathsf P})^{\otimes m}\middle\Vert
   (\Omega_{1,n}^{\mathsf P})^{\otimes m}\right)
 =D(\Omega_{0,n}^{\mathsf P}\Vert\Omega_{1,n}^{\mathsf P})
\end{equation}
for every fixed preparation and $0<\varepsilon<1$
\cite{HiaiPetz1991,OgawaNagaoka2000}. Consequently, $I_n^D$ is the
minimum of the known-device Stein exponents, with the minimum taken
\emph{after} the limit for each device. In contrast, the universal test
in \eqref{eq:beta-opt} must be chosen before the device is specified.
Its rate is therefore bounded above by $I_n^D$, but the ordinary Stein
lemma alone does not give a matching universal test.

The following theorem establishes the universal exponent and compares
it with the preparation cost. The quantitative estimate uses the same
measurement as Lemma~\ref{lem:quantitative-horizon}: it extracts from
every compatible output pair classical distributions whose relative
entropy is close to the BS value. The common choice of measurement is
essential for a statement that is uniform over preparations.

\begin{theorem}[Universal Stein exponent for $n$-use preparations]
\label{lem:fixed-block}
Let $\rho_0,\rho_1>0$ be states on $S$.
For every $n\ge1$ and $0<\varepsilon<1$, the limit
\begin{equation}\label{eq:rn}
r_n(\rho_0\Vert\rho_1)
:=\lim_{m\to\infty}-\frac1m\log\beta_{n,m}^{\varepsilon}
\end{equation}
exists and is independent of $\varepsilon$.
The sequence $(r_n)_{n\ge1}$ is nondecreasing and satisfies
\begin{equation}\label{eq:rn-basic-bounds}
D(\rho_0\Vert\rho_1)
=r_1(\rho_0\Vert\rho_1)
\le r_n(\rho_0\Vert\rho_1)
\le I_n^D(\rho_0\Vert\rho_1)
\le\BS(\rho_0\Vert\rho_1).
\end{equation}
Moreover, for $d=\dim S$ and  for
\[
 \lambda=\min\{\lambda_{\min}(\rho_0),\lambda_{\min}(\rho_1)\}\, , \qquad L_{\rho_0,\rho_1}=\log\frac{2}{\lambda}
             +\frac{2\sqrt{2}}{\lambda^{5/2}}\, ,
 \qquad n_0=\left\lceil\frac{4d^2}{\lambda}\right\rceil.
\]
the lower bound improves to
\begin{equation}\label{eq:rn-bounds}
 r_n(\rho_0\Vert\rho_1)
 \ge\max\left\{
 D(\rho_0\Vert\rho_1),
 \BS(\rho_0\Vert\rho_1)-\frac{4d^2L_{\rho_0,\rho_1}}{n}
 \right\}.
\end{equation}
for every $n\geq n_0$.
\end{theorem}
\begin{proof}
The proof is given in Theorem~\ref{AP-lem:fixed-block}.
\end{proof}

\subsection{The Belavkin--Staszewski Stein limit}
\label{subsec:universal-bs-limit}

Theorem~\ref{lem:fixed-block} gives the universal testing exponent when
$n$, the number of emissions per initialization, is fixed. Its estimates
already show that the cost of not knowing the preparation vanishes as
$n$ increases. More precisely, for the constants $L_{\rho_0,\rho_1}$
and $n_0$ of that theorem,
\begin{equation}\label{eq:preparation-testing-gap}
 0\le I_n^D(\rho_0\Vert\rho_1)-r_n(\rho_0\Vert\rho_1)
 \le\frac{4d^2L_{\rho_0,\rho_1}}{n},\qquad n\ge n_0.
\end{equation}
Thus the optimal known-device and universal rates have the same limit,
which Theorem~\ref{thm:preparation} identifies with the information cost of
stationary preparation.

A stronger operational question arises when the number of emissions in
each round is itself allowed to increase with the number of independent
initializations. The fixed-$n$ limit alone does not settle this question,
since convergence in the number of rounds need not be uniform in $n$.
Nevertheless, any admissible preparation can be truncated to fewer
emissions, and a single stationary classical preparation provides a
converse simultaneously for every block length.

\begin{theorem}[Belavkin--Staszewski Stein limit]
\label{thm:BS-Stein}
Let $\rho_0,\rho_1>0$ be states on $S$, and let
$0<\varepsilon<1$. The universal exponents of
Theorem~\ref{lem:fixed-block} satisfy
\begin{equation}\label{eq:iterated-Stein}
 \lim_{n\to\infty}r_n(\rho_0\Vert\rho_1)
 =\BS(\rho_0\Vert\rho_1).
\end{equation}
More generally, for every sequence of positive integers
$n_m\to\infty$,
\begin{equation}\label{eq:simultaneous-Stein}
 \lim_{m\to\infty}-\frac1m\log\beta_{n_m,m}^{\varepsilon}
 =\BS(\rho_0\Vert\rho_1).
\end{equation}
No restriction on the growth of $n_m$ relative to $m$ is required.
\end{theorem}

\begin{proof}
The proof is given in Theorem~\ref{AP-thm:BS-Stein}.
\end{proof}

\section*{Acknowledgments}
 Á.C. acknowledges support from the Deutsche Forschungsgemeinschaft
(DFG, German Research Foundation), Project-ID 470903074, TRR 352.
This project was funded within the QuantERA II Programme, which
received funding from the European Union's Horizon 2020 research and
innovation programme under Grant Agreement No. 101017733. PCR, MB and RS acknowledge support from the European
Research Council (ERC Grant Agreement No.\ 948139) and from the Excellence Cluster --
Matter and Light for Quantum Computing (ML4Q-2).

\paragraph{AI Statement:}
The authors acknowledge the use of  ChatGPT 6 Astra to assist with brainstorming, drafting and organizing the literature review. The final output was thoroughly verified by the authors, who take full responsibility for the accuracy and integrity of this work.

\bibliographystyle{abbrv}
\bibliography{ReferencesQPWD}

\clearpage
\appendix

\section{Proofs}

\subsection{Stationary preparation and memory testing}
\label{app:stationary-proofs}

\begin{theorem}[Stationary Preparation]
\label{AP-thm:preparation}
Let $
\rho_0,\rho_1\in\mathcal D(S)$ satisfying $
\operatorname{supp}\rho_0\subseteq\operatorname{supp}\rho_1
$, and let $\mathbb{D}$ be a quantum divergence, with classical restriction $\mathbb{D}_{\mathrm{cl}}$. Then
\begin{equation}
  \inf_{\substack{M,\omega_0,\omega_1,\cR:M\to M\otimes S\\
                  \tr_M\cR(\omega_i)=\rho_i\ (i=0,1)}}
       \mathbb{D}(\omega_0\|\omega_1)
  =\mathbb{D}(\rho_0\|\rho_1).
  \label{AP-eq:unrestricted-minimum}
\end{equation}
With the additional memory-preservation conditions,
\begin{equation}
  \inf_{\substack{M,\omega_0,\omega_1,\cR:M\to M\otimes S\\
                  \tr_S\cR(\omega_i)=\omega_i\\
                  \tr_M\cR(\omega_i)=\rho_i\ (i=0,1)}}
       \mathbb{D}(\omega_0\|\omega_1)
  =\widehat{\mathbb{D}}(\rho_0\|\rho_1).
  \label{AP-eq:stationary-minimum}
\end{equation}
The infimum in \eqref{AP-eq:unrestricted-minimum} is attained. In
\eqref{AP-eq:stationary-minimum}, restricting to classical memories does not
change the infimum. If the reverse-preparation infimum
\eqref{eq:maximal-extension} is attained, the corresponding classical memory
attains \eqref{eq:stationary-minimum}.
\end{theorem}

\begin{proof}
For the unrestricted preparation problem, the output channel
$\mathcal{O}:=\tr_M\circ\cR$ satisfies $\mathcal{O}(\omega_i)=\rho_i$.
Data processing therefore implies
$\mathbb D(\omega_0\|\omega_1)\ge\mathbb D(\rho_0\|\rho_1)$.
Conversely, take $M\simeq S$, set $\omega_i=\rho_i$, and define
$\cR(X)=\chi\otimes X$ for an arbitrary state $\chi\in\D(M)$.
This channel satisfies $\cR(\omega_i)=\chi \otimes \omega_i=\chi \otimes \rho_i$ and therefore $\mathcal{O}(\omega_i)=\rho_i$ proving \eqref{AP-eq:unrestricted-minimum} and its
attainment.

We turn to the stationary minimum. Fix an admissible quadruple
$(M,\omega_0,\omega_1,\cR)$, and denote by
$\mathcal T:=\tr_S\circ\cR$ its retained-memory channel. The
stationarity assumption says that $\mathcal T(\omega_i)=\omega_i$
for $i=0,1$. Choose a Stinespring isometry
$V:M\to M\otimes S\otimes E$ such that
\[
 \cR(X)=\tr_E(VXV^*),\qquad
 \mathcal T(X)=\tr_{S\otimes E}(VXV^*).
\]
In particular, $V$ is a Stinespring isometry for $\mathcal T$ when
$S\otimes E$ is regarded as its environment. The preservation of the
two memory states is thus a  condition to which the
Koashi--Imoto structure theorem applies.

The justification that the memory states can be restricted to their joint support with projection $P$ can be found in Lemma \ref{lemma:SupportRestriction}. This reduction is necessary because the
admissible records are not assumed to be faithful. 

We now invoke the Koashi--Imoto  theorem
\cite[Theorem~3]{KoashiImoto2002} for the family
$\{\omega_0,\omega_1\}$ and the isometry $V$ on $M_P=PM$. It provides
a direct-sum decomposition
\begin{equation}
 M_P=\bigoplus_z M_z^L\otimes M_z^R,
 \qquad
 \omega_i=\bigoplus_z p_i(z)\,a_{i,z}\otimes b_z,
 \label{eq:ki-decomposition}
\end{equation}
where $p_i$ are probability distributions, $a_{i,z}$ and $b_z$
are normalized states, and $b_z$ is independent of $i$.  The theorem further asserts that every isometry
preserving both memory states has, under this decomposition, the form
\[
 V\big|_{M_P}=\bigoplus_z I_{M_z^L}\otimes V_z,
 \qquad
 V_z:M_z^R\longrightarrow M_z^R\otimes S\otimes E,
\]
with isometries $V_z$ independent of $i$. 

Define
$\Gamma_z:=\tr_E(V_zb_zV_z^*)\in\D(M_z^R\otimes S)$.
Because both $V_z$ and $b_z$ are independent of $i$, so is
$\Gamma_z$. Applying the isometry to the states in
\eqref{eq:ki-decomposition} gives
\begin{equation}
 \cR(\omega_i)
   =\bigoplus_z p_i(z)\,a_{i,z}\otimes\Gamma_z.
 \label{eq:ki-output}
\end{equation}
Moreover, $\tr_S\Gamma_z=b_z$ and for each $z$, at least one
of $p_0(z)$ and $p_1(z)$ is nonzero because $M_P$ is their joint
support. Finally, put
$\tau_z:=\tr_{M_z^R}\Gamma_z\in\D(S)$.
Tracing \eqref{eq:ki-output} over the memory yields
\begin{equation}
 \rho_i=\tr_M\cR(\omega_i)=\sum_zp_i(z)\tau_z,
 \qquad i=0,1.
 \label{eq:extracted-reverse-preparation}
\end{equation}

Let $P_z$ be the projection onto $M_z^L\otimes M_z^R$.
The projective measurement
$\mathcal M(X)=\sum_z\tr(P_zX)|z\rangle\langle z|$ on $M_P$
satisfies $\mathcal M(\omega_i)=\diag p_i$.
Using data processing, the classical restriction of $\mathbb D$,
and the reverse preparation just constructed, we obtain
\begin{equation}\label{eq:stationary-lower-bound}
 \mathbb D(\omega_0\|\omega_1)
 \ge\mathbb D_{\mathrm{cl}}(p_0\|p_1)
 \ge\widehat{\mathbb D}(\rho_0\|\rho_1)\, .
\end{equation}
The second inequality is precisely the infimum property in
\eqref{eq:maximal-extension}. Since the stationary preparation was
arbitrary, this proves the lower bound in
\eqref{eq:stationary-minimum}.

For the converse, take an arbitrary finite classical reverse
preparation $\rho_i=\sum_zp_i(z)\tau_z$, and let the memory have an
orthonormal basis $\{|z\rangle\}_z$. Define
\begin{equation}
 \omega_i^{\mathrm{cl}}=\sum_zp_i(z)|z\rangle\langle z|,
 \qquad
 \cR_{\mathrm{cl}}(X)
 =\sum_z\langle z|X|z\rangle
       |z\rangle\langle z|\otimes\tau_z.
 \label{eq:classical-stationary-device}
\end{equation}
This is a common channel for both inputs, and direct partial traces
give
\[
 \tr_M\cR_{\mathrm{cl}}(\omega_i^{\mathrm{cl}})=\rho_i,
 \qquad
 \tr_S\cR_{\mathrm{cl}}(\omega_i^{\mathrm{cl}})
    =\omega_i^{\mathrm{cl}}.
\]
Therefore, the quadruple $(M,\omega_0^{\mathrm{cl}},\omega_1^{\mathrm{cl}},\cR_{\mathrm{cl}})$ is a stationary preparation and by the classical restriction property,
 \[ \inf_{\substack{M,\omega_0,\omega_1,\cR:M\to M\otimes S\\
                  \tr_S\cR(\omega_i)=\omega_i\\
                  \tr_M\cR(\omega_i)=\rho_i\ (i=0,1)}}
       \mathbb{D}(\omega_0\|\omega_1)
 \leq \mathbb{D}(\omega_0^{\mathrm{cl}}\Vert \omega_1^{\mathrm{cl}})=\mathbb D_{\mathrm{cl}}(p_0\|p_1).\]
Taking the infimum over all reverse preparations yields the
opposite inequality in \eqref{eq:stationary-minimum}. Together with
\eqref{eq:stationary-lower-bound}, this proves the claimed equality.
The same construction shows that classical memories suffice, and it
attains the stationary infimum whenever the classical
reverse-preparation infimum is attained.
\end{proof}

\begin{theorem}[Robustness of stationary preparation]
\label{AP-thm:stationary-robustness}
Let $\rho_0,\rho_1>0$ be states on $S$, and let $\mathbb D$ satisfy
data processing with classical restriction
$\mathbb D_{\mathrm{cl}}=D_{\mathrm{KL}}$. Write $d=\dim S$ and set
\[
 \lambda=\min\{\lambda_{\min}(\rho_0),\lambda_{\min}(\rho_1)\},
 \qquad
 L_{\rho_0,\rho_1}
 =\log\frac{2}{\lambda}+\frac{2\sqrt{2}}{\lambda^{5/2}}.
\]
For $\varepsilon_{\mathrm s},\varepsilon_{\mathrm o}\geq0$ satisfying
$\varepsilon_{\mathrm o}+2d\sqrt{\varepsilon_{\mathrm s}}
\leq\lambda/2$, the cost in
\eqref{eq:approximate-stationary-cost} obeys
\begin{equation}\label{AP-eq:stationary-robust-bound}
 0\leq\BS(\rho_0\Vert\rho_1)
 -I_{\varepsilon_{\mathrm s},\varepsilon_{\mathrm o}}^{\mathbb D}
 (\rho_0\Vert\rho_1)
 \leq 2L_{\rho_0,\rho_1}
 \bigl(\varepsilon_{\mathrm o}
       +2d\sqrt{\varepsilon_{\mathrm s}}\bigr).
\end{equation}
In particular,
\begin{equation}\label{AP-eq:stationary-robust-limit}
 \lim_{
 \varepsilon_{\mathrm s},\varepsilon_{\mathrm o}\to0^+}
 I_{\varepsilon_{\mathrm s},\varepsilon_{\mathrm o}}^{\mathbb D}
 (\rho_0\Vert\rho_1)
 =\BS(\rho_0\Vert\rho_1).
\end{equation}
\end{theorem}

\begin{proof}

For $\varepsilon_{\mathrm s},\varepsilon_{\mathrm o}\geq 0$, enlarging 
the set of admissible preparations cannot
increase the infimum  information cost. Hence, for every
$\varepsilon_{\mathrm s},\varepsilon_{\mathrm o}\geq 0$,
Theorem~\ref{thm:preparation} gives
\begin{equation}\label{eq:robust-exact-upper-bound}
 I_{\varepsilon_{\mathrm s},\varepsilon_{\mathrm o}}^{\mathbb D}
 (\rho_0\Vert\rho_1)
 \leq I_{0,0}^{\mathbb D}(\rho_0\Vert\rho_1)
 =\BS(\rho_0\Vert\rho_1).
\end{equation}

For the opposite estimate, fix an arbitrary admissible preparation
$(M,\omega_0,\omega_1,\cR)$ and write
$\mathcal T=\tr_S\circ\cR$ and
$\mathcal O=\tr_M\circ\cR$. We first control the accumulated error under repeated use. For each
$j\geq0$, a telescoping sum and trace-norm contractivity give
\[
 \|\mathcal T^j(\omega_i)-\omega_i\|_1 \leq\sum_{k=0}^{j-1}
 \bigl\|\mathcal T^k\bigl(\mathcal T(\omega_i)-\omega_i\bigr)\bigr\|_1
 \leq j\varepsilon_{\mathrm s}.
\]
Consequently,
\begin{equation}\label{eq:robust-output-drift}
 \|\mathcal O(\mathcal T^j(\omega_i))-\rho_i\|_1
 \leq 
\|\mathcal O(\mathcal T^j(\omega_i))
      -\mathcal O(\omega_i)\|_1
+\|\mathcal O(\omega_i)-\rho_i\|_1 \leq j\varepsilon_{\mathrm s}+\varepsilon_{\mathrm o},
 \qquad i=0,1.
\end{equation}

For an integer $n\geq1$, let
$\Psi_n=\tr_M\circ\cR^{[n]}:M\to S^{\otimes n}$
collect the first $n$ emissions, as in
\eqref{eq:sequential-output-channel}.
By the marginal identities
\eqref{eq:sequential-output-marginals}, the average-marginal
channel $\mathcal N_n$ from \eqref{eq:average-marginal}
satisfies
\[
\mathcal N_n(\Psi_n(\omega_i))
=\frac1n\sum_{j=0}^{n-1}
\mathcal O(\mathcal T^j(\omega_i)).
\]
Consequently, the triangle inequality and
\eqref{eq:robust-output-drift} give
\[
\begin{aligned}
\|\mathcal N_n(\Psi_n(\omega_i))-\rho_i\|_1
&\leq\frac1n\sum_{j=0}^{n-1}
\|\mathcal O(\mathcal T^j(\omega_i))-\rho_i\|_1\\
&\leq\frac1n\sum_{j=0}^{n-1}
(j\varepsilon_{\mathrm s}+\varepsilon_{\mathrm o})\\
&=\varepsilon_{\mathrm o}
+\frac{n-1}{2}\varepsilon_{\mathrm s}.
\end{aligned}
\]

We now apply Lemma~\ref{lem:extraction} to the collected outputs.
It provides a common measure-and-prepare channel
\[
 \Phi_n(X)=\sum_z\tr(F_z^{(n)}X)\tau_z^{(n)}
\]
whose trace-norm distance from $\mathcal N_n$ on every input state is
at most $2d^2/n$. For
$a_i^{(n)}:=\Phi_n(\Psi_n(\omega_i))$, the preceding estimates yield
\begin{equation}\label{eq:robust-extracted-error}
 \|a_i^{(n)}-\rho_i\|_1
 \leq\varepsilon_{\mathrm o}
     +\frac{n-1}{2}\varepsilon_{\mathrm s}
     +\frac{2d^2}{n}
 =:\eta_n,
 \qquad i=0,1.
\end{equation}
Because $\Psi_n$ is a channel, the operators
$E_z^{(n)}:=\Psi_n^*(F_z^{(n)})$ form a POVM on $M$.
Writing
\[
 p_{i,n}(z)=\tr(E_z^{(n)}\omega_i),
 \qquad
 a_i^{(n)}=\sum_zp_{i,n}(z)\tau_z^{(n)},
\]
we obtain a common classical reverse preparation of
$a_0^{(n)},a_1^{(n)}$.
Data processing under this measurement, followed by the
reverse-preparation characterization of $\BS$, gives
\begin{equation}\label{eq:robust-extracted-cost}
 \mathbb D(\omega_0\Vert\omega_1)
 \geq\KL(p_{0,n}\Vert p_{1,n})
 \geq\BS(a_0^{(n)}\Vert a_1^{(n)}).
\end{equation}

Whenever $\eta_n\leq\lambda/2$, the extracted states satisfy
\[
 a_i^{(n)}
 \geq\rho_i-\|a_i^{(n)}-\rho_i\|_\infty I
 \geq(\lambda-\eta_n)I
 \geq\frac{\lambda}{2}I.
\]
We may therefore use the BS continuity estimate
\eqref{eq:BS-explicit-Lipschitz} established in the proof of
Lemma~\ref{lem:quantitative-horizon}, with the constant
$L_{\rho_0,\rho_1}$ stated above. Together with
\eqref{eq:robust-extracted-error} and
\eqref{eq:robust-extracted-cost}, it gives
\[
 \mathbb D(\omega_0\Vert\omega_1)
 \geq\BS(\rho_0\Vert\rho_1)
       -L_{\rho_0,\rho_1}
        \sum_{i=0}^1\|a_i^{(n)}-\rho_i\|_1
 \geq\BS(\rho_0\Vert\rho_1)-2L_{\rho_0,\rho_1}\eta_n.
\]
This bound holds for every admissible preparation, independently of
its memory dimension. Taking the infimum thus yields
\begin{equation}\label{eq:robust-cost-at-n}
 I_{\varepsilon_{\mathrm s},\varepsilon_{\mathrm o}}^{\mathbb D}
 (\rho_0\Vert\rho_1)
 \geq\BS(\rho_0\Vert\rho_1)-2L_{\rho_0,\rho_1}\eta_n
 \qquad\text{if }\eta_n\leq\lambda/2.
\end{equation}

For $\varepsilon_{\mathrm s}>0$, choose
\[
 n=\left\lceil\frac{2d}{\sqrt{\varepsilon_{\mathrm s}}}\right\rceil.
\]
Then
\[
 \frac{n-1}{2}\varepsilon_{\mathrm s}
 \leq d\sqrt{\varepsilon_{\mathrm s}},
 \qquad
 \frac{2d^2}{n}\leq d\sqrt{\varepsilon_{\mathrm s}},
\]
and hence
$\eta_n\leq\varepsilon_{\mathrm o}
+2d\sqrt{\varepsilon_{\mathrm s}}\leq\lambda/2$.
Substitution in \eqref{eq:robust-cost-at-n}, together with
\eqref{eq:robust-exact-upper-bound}, proves
\eqref{AP-eq:stationary-robust-bound}.

If $\varepsilon_{\mathrm s}=0$, the memory is exactly stationary.
Its actual emitted states $a_i=\mathcal O(\omega_i)$ satisfy
$\|a_i-\rho_i\|_1\leq\varepsilon_{\mathrm o}\leq\lambda/2$,
and thus $a_i\geq(\lambda/2)I$.
Applying Theorem~\ref{thm:preparation} to the pair $(a_0,a_1)$,
followed by the same continuity estimate, gives
\[
 \mathbb D(\omega_0\Vert\omega_1)
 \geq\BS(a_0\Vert a_1)
 \geq\BS(\rho_0\Vert\rho_1)
       -2L_{\rho_0,\rho_1}\varepsilon_{\mathrm o}.
\]
Taking the infimum proves
\eqref{AP-eq:stationary-robust-bound} in this case as well.
Finally, the right-hand side of that bound tends to zero as both
errors vanish, proving \eqref{AP-eq:stationary-robust-limit}.
\end{proof}

\begin{theorem}[Stein limit for stationary memories]\label{AP-thm:source-Stein}
Let $\rho_0,\rho_1\in\D(S)$ satisfy
$\supp\rho_0\subseteq\supp\rho_1$. Then, for every fixed
$0<\varepsilon<1$,
\begin{equation}\label{AP-eq:source-Stein}
 \lim_{n\to\infty}\frac1n
 D_{H}^{\varepsilon,n}(\rho_0\Vert\rho_1)
 =\BS(\rho_0\Vert\rho_1).
\end{equation}
\end{theorem}

\begin{proof}
Since $\supp\rho_0\subseteq\supp\rho_1$, we may restrict the
output space to $\supp\rho_1$ and assume that $\rho_1$ is faithful.
The state $\rho_0$ need not be faithful.

For $0<\alpha<1$, the standard  R\'enyi testing bound
 \cite[Lemma~3.2]{BertaBrandaoHirche2021}
gives, for any probability distributions $p,q$,
\[
 D_H^\varepsilon(p^{\otimes n}\|q^{\otimes n})
 \geq nD_\alpha(p\|q)
       -\frac{\alpha}{1-\alpha}\log\frac1\varepsilon.
\]
Taking the infimum over classical preparations, we obtain
\begin{equation}\label{eq:source-uniform-renyi}
 \frac1nD_{H}^{\varepsilon,n}(\rho_0\|\rho_1)
 \geq \widehat D_\alpha(\rho_0\|\rho_1)
       -\frac{\alpha}{n(1-\alpha)}\log\frac1\varepsilon.
\end{equation}

Let $K=\rho_1^{-1/2}\rho_0\rho_1^{-1/2}\geq0$, with the inverse
taken on $\supp\rho_1$. The geometric R\'enyi divergence can be expressed as
\[
 \widehat D_\alpha(\rho_0\|\rho_1)
 =\frac{1}{\alpha-1}\log\tr(\rho_1 K^\alpha),
 \qquad 0<\alpha<1.
\]
Since $\tr(\rho_1K)=1$, differentiation from the left at $\alpha=1$
yields
\[
 \lim_{\alpha\uparrow1}\widehat D_\alpha(\rho_0\|\rho_1)
 =\tr(\rho_1K\log K)=\BS(\rho_0\|\rho_1),
\]
 Letting first $n\to\infty$ and then
$\alpha\uparrow1$ in \eqref{eq:source-uniform-renyi} proves the
required lower bound.

For the upper bound, use the optimal classical reverse preparation
associated with the spectral decomposition 
$K=\sum_z\lambda_zP_z$. It has
$q(z)=\tr(\rho_1P_z)>0$ and $p(z)=\lambda_zq(z)$, and satisfies
\[
 D_{\mathrm{KL}}(p\|q)
 =\tr(\rho_1K\log K)=\BS(\rho_0\|\rho_1).
\]
This reverse preparation defines a fixed stationary classical source
by \eqref{eq:classical-stationary-device}. The classical Stein lemma
\cite{CsiszarKorner2011} then implies
\[
 \limsup_{n\to\infty}\frac1n
 D_H^{\varepsilon,n}(\rho_0\|\rho_1)
 \leq \lim_{n\to\infty}\frac1n
 D_H^\varepsilon(p^{\otimes n}\|q^{\otimes n})
 =D_{\mathrm{KL}}(p\|q)=\BS(\rho_0\|\rho_1).
\]
Together with the lower bound, this proves \eqref{AP-eq:source-Stein}.
\end{proof}

\subsection{Repeated preparation and the divergence hierarchy}

\begin{proposition}[Infinite memory chain]
\label{AP-prop:output-stationarity}\label{AP-cor:output-stationarity}
Let $\rho_0,\rho_1\in\D(S)$ satisfy
$\supp\rho_0\subseteq\supp\rho_1$, and let $\mathbb D$
be a quantum divergence. Then,
\begin{equation}\label{AP-eq:output-stationarity-minimum}
 \inf_{\substack{M,\omega_0,\omega_1,\cR:M\to M\otimes S\\
                  \mathcal O(\mathcal T^k(\omega_i))=\rho_i\\
                  i=0,1,\ k\ge0}}
       \mathbb D(\omega_0\Vert\omega_1)\\
=\widehat{\mathbb D}(\rho_0\Vert\rho_1)\, .
\end{equation}
\end{proposition}

\begin{proof}
Every stationary preparation satisfies the output conditions for all uses, since $\mathcal{T}^k(\omega_i)=\omega_i$ implies $\mathcal{O}(\mathcal{T}^k(\omega_i))=\rho_i$, for every $k\geq 0$. Thus, stationary preparations form a subset of infinite use preparations, which shows that
\begin{equation}
     \inf_{\substack{M,\omega_0,\omega_1,\cR:M\to M\otimes S\\
                  \mathcal O(\mathcal T^k(\omega_i))=\rho_i\\
                  i=0,1,\ k\ge0}}
       \mathbb D(\omega_0\Vert\omega_1)\\
\leq \widehat{\mathbb D}(\rho_0\Vert\rho_1)\, .
\end{equation}

To prove the reverse inequality, fix a device admissible
for indefinitely many uses and consider the Ces\`aro averages
\[
 \mathcal P_N=\frac1N\sum_{k=0}^{N-1}\mathcal T^k.
\]
Each $\mathcal P_N$ is a channel on the same finite-dimensional memory. The set of channels on this memory is closed and bounded in a finite-dimensional space, hence compact. Therefore, there exists a subsequence converging to a channel $\mathcal P$. Moreover,
\[
 \mathcal T\circ\mathcal P_N-\mathcal P_N
   =\frac{\mathcal T^N-\id}{N},
\]
which converges to zero, since the powers of a channel are uniformly
bounded, namely
\[
    \Vert \mathcal T\circ\mathcal P_N-\mathcal P_N\Vert_{1 \to 1}
   =\frac{1}{N}\Vert \mathcal T^N-\id \Vert_{1\to 1}\leq \frac{2}{N}\xrightarrow{N \to \infty}0 \, .
\]
 Hence $\mathcal T\circ\mathcal P=\mathcal P$. For
$\widetilde\omega_i=\mathcal P(\omega_i)$, linearity of $\mathcal O$ and the
output conditions give
\[
 \mathcal O(\widetilde\omega_i)
   =\lim_{j\to\infty}\frac1{N_j}\sum_{k=0}^{N_j-1}
           \mathcal O(\mathcal T^k(\omega_i))=\rho_i,
 \qquad
 \mathcal T(\widetilde\omega_i)=\widetilde\omega_i.
\]

Thus the quadruple $(M,\widetilde\omega_0,\widetilde\omega_1, \cR)$ is a stationary preparation. Data processing under the  channel
$\mathcal P$ yields
\begin{equation}\label{eq:Inequality-chain-preparations}
 \mathbb D(\omega_0\Vert\omega_1) \ge \mathbb D(\widetilde\omega_0\Vert\widetilde\omega_1) \ge \inf_{\substack{M, \widetilde \omega_0,\widetilde \omega_1,\cR:M\to M\otimes S\\
                  \tr_S\cR(\widetilde \omega_i)=\widetilde \omega_i\\
                  \tr_M\cR(\widetilde \omega_i)=\rho_i\ (i=0,1)}}
       \mathbb{D}(\widetilde\omega_0\|\widetilde\omega_1)=\widehat{\mathbb{D}}(\rho_0\Vert \rho_1).
\end{equation}
where the last equality follows by
Theorem~\ref{thm:preparation}. Since this holds for every infinite preparation, taking the infimum over such preparations on the left-hand side of \eqref{eq:Inequality-chain-preparations} gives the reverse inequality.
\end{proof}

\begin{theorem}[Data processing and classical restriction]
\label{AP-thm:horizon-divergences}
Let $\mathbb D$ be a quantum divergence with classical
restriction $\mathbb{D}_{\mathrm{cl}}$, and let
$\rho_0,\rho_1\in\D(S)$ satisfy
$\supp\rho_0\subseteq\supp\rho_1$. For every
$n\in\mathbb N\cup\{+\infty\}$ and every channel $\Phi:S\to S'$, one has
\begin{equation}\label{AP-eq:horizon-dpi}
 I_n^{\mathbb D}\bigl(\Phi(\rho_0)\Vert\Phi(\rho_1)\bigr)
 \le I_n^{\mathbb D}(\rho_0\Vert\rho_1).
\end{equation}
Moreover, for probability distributions $p,q$ with
$\supp p\subseteq\supp q$,
\begin{equation}\label{AP-eq:horizon-classical-restriction}
 I_n^{\mathbb D}(\diag p\Vert\diag q)
   =\mathbb D_{\mathrm{cl}}(p\Vert q).
\end{equation}
\end{theorem}

\begin{proof}
The support inclusion is preserved by $\Phi$: in finite dimensions the constraint $\supp\rho_0\subseteq\supp\rho_1$ implies the existence of a positive constant $c$ such that 
$\rho_0\le c\rho_1$, and positivity and linearity of  $\Phi$ gives
$\Phi(\rho_0)\le c\Phi(\rho_1)$. Hence $\supp\Phi(\rho_0)\subseteq\supp\Phi(\rho_1)$.

We prove data processing directly from the
preparation model. Fix a quadruple $(M,\omega_0,\omega_1,\cR)$ admissible for the $n$-use preparation cost,
and postprocess each emitted system by $\Phi$. The resulting device has
channel
\[
 \cR_\Phi=(\id_M\otimes\Phi)\circ\cR:M\longrightarrow M\otimes S'.
\]
Trace preservation of $\Phi$ leaves the memory evolution unchanged,
whereas its output channel is composed with $\Phi$:
\[
 \mathcal T_\Phi=\tr_{S'}\circ\cR_\Phi=\mathcal T,
 \qquad
 \mathcal O_\Phi=\tr_M\circ\cR_\Phi=\Phi\circ\mathcal O.
\]
For each required use we  have then
\[
 \mathcal O_\Phi\bigl(\mathcal T_\Phi^k(\omega_i)\bigr)
   =\Phi\bigl(\mathcal O(\mathcal T^k(\omega_i))\bigr)
   =\Phi(\rho_i).
\]
Therefore, $(M,\omega_0,\omega_1,\mathcal R_\Phi)$ is an admissible preparation of \(\Phi(\rho_0),\Phi(\rho_1)\) for exactly the same number of uses. In particular, $I_n^{\mathbb{D}}\bigl(\Phi(\rho_0)\Vert\Phi(\rho_1)\bigr)\le \mathbb{D}(\omega_0\Vert \omega_1)$. Since this last inequality holds for every admissible preparation of the original states $\rho_0$ and $\rho_1$, we can take the infimum over them:
\[
I_n^{\mathbb D}
\bigl(\Phi(\rho_0)\Vert\Phi(\rho_1)\bigr)
\le  \inf_{\substack{M,\omega_0,\omega_1,\cR:M\to M\otimes S\\
       \mathcal O(\mathcal T^k(\omega_i))=\rho_i\\
       i=0,1,\quad k=0,\ldots,n-1}}
       \mathbb D(\omega_0\Vert\omega_1)
=I_n^{\mathbb D}(\rho_0\Vert\rho_1).
\]
The same argument follows for $n=\infty$.

To show the second part, let $(M,\omega_0,\omega_1,\mathcal{R})$ be an admissible $n$-use preparation, $n\geq 1$.  By  data processing inequality we get
\[
 \mathbb{D}(\omega_0\Vert \omega_1)\ge \mathbb{D}(\mathcal{O}(\omega_0)\Vert \mathcal{O} (\omega_1))
  =\mathbb D(\rho_0\Vert\rho_1).
\]
Taking the infimum over all admissible preparations
therefore gives $
I_n^{\mathbb D}(\rho_0\Vert\rho_1)
\geq\mathbb D(\rho_0\Vert\rho_1)$. 
In particular, when the target states are diagonal, $
\rho_i=\operatorname{diag}p_i,$ the classical restriction of \(\mathbb D\) gives
\[
I_n^{\mathbb D}(\operatorname{diag}p_0
\Vert\operatorname{diag}p_1)
\ge
\mathbb D_{\mathrm{cl}}(p_0\Vert p_1).
\]

On diagonal targets, the opposite bound is realized by the classical
records $\omega_0=\diag p$, $\omega_1=\diag q$ and the channel
\[
 X\longmapsto\sum_z\langle z|X|z\rangle
        |z\rangle\langle z|\otimes|z\rangle\langle z|.
\]
It preserves both records and gives the required marginals at every use,
with cost $\mathbb D_{\mathrm{cl}}(p\Vert q)$. This proves
\eqref{eq:horizon-classical-restriction}.
\end{proof}

\begin{proposition}[Parallel preparation]
\label{AP-prop:parallel-preparation}
Let $n\ge1$, let $\rho_0,\rho_1\in\D(S)$ with $\operatorname{supp}\rho_0\subseteq\operatorname{supp}\rho_1$, and let $\mathbb D$ satisfy
data processing. Then
\begin{equation}\label{AP-eq:finite-use}
 I_n^{\mathbb D}(\rho_0\Vert\rho_1)
  =\inf_{\substack{M,\omega_0,\omega_1,\Psi:M\to S^{\otimes n}\\
          [\Psi(\omega_i)]_{S_j}=\rho_i\\
          i=0,1,\ 1\le j\le n}}
        \mathbb D(\omega_0\Vert\omega_1),
\end{equation}
and
\begin{equation}\label{AP-eq:horizon-extension-equivalence}
 I_n^{\mathbb D}(\rho_0\Vert\rho_1)
  =\inf_{\substack{R_n\in\cE_n(\rho_0)\\T_n\in\cE_n(\rho_1)}}
       \mathbb D(R_n\Vert T_n).
\end{equation}
Moreover, if \(\mathbb D\) is jointly lower semicontinuous and its classical restriction is finite for all pairs of probability distributions with support inclusion, then, under \(\operatorname{supp}\rho_0\subseteq\operatorname{supp}\rho_1\), all three infima are finite and attained. A memory of dimension \((\dim S)^n\) suffices.
\end{proposition}

\begin{proof}
Write $J_n$ for the parallel infimum on the right side of
\eqref{eq:finite-use}, and $E_n$ for the extension infimum on the right
side of \eqref{eq:horizon-extension-equivalence}. We prove
$E_n\le J_n\le I_n^{\mathbb D}\le E_n$ by explicit channel constructions.

The inequality $J_n\le I_n^{\mathbb D}$ is clear by Definition \ref{def:I_n}, since every admissible sequential device defines the common parallel channel
$\Psi_n=\tr_M\circ\cR^{[n]}$ from
\eqref{eq:sequential-output-channel}. Its marginals are given by
\eqref{eq:sequential-output-marginals}, so it satisfies
\eqref{eq:parallel-marginal-condition} with the same input records and
the same cost.

Conversely, take any parallel preparation and set
$R_n=\Psi(\omega_0)$, $T_n=\Psi(\omega_1)$. These states belong to the
respective extension sets, and data processing gives
\[
 E_n\le\mathbb D(R_n\Vert T_n)
       \le\mathbb D(\omega_0\Vert\omega_1).
\]
Taking the infimum over parallel devices yields $E_n\le J_n$.

It remains to show that every pair of compatible extensions
can also be realized by a sequential preparation. Fix
$R_n\in\cE_n(\rho_0)$ and $T_n\in\cE_n(\rho_1)$. Let  $M=A_1\otimes\cdots\otimes A_n$,
where $A_j\simeq S$, and choose the initial records
$\omega_0=R_n$ and $\omega_1=T_n$.
We will construct a single preparation channel whose
first $n$ outputs reproduce these states exactly.

Fix an arbitrary state $\chi\in\D(S)$, which will be used
to replace each register after it has been emitted.
Consider the tensor-factor permutation
\[
 V(v_1\otimes\cdots\otimes v_n\otimes w)
 =v_2\otimes\cdots\otimes v_n\otimes w\otimes v_1,
\]
where the first $n$ factors on the right correspond to
the updated memory and the last factor to the emitted
system. This defines the preparation channel $
\mathcal R_\chi:\mathcal B(S^{\otimes n})
\longrightarrow \mathcal B(S^{\otimes(n+1)})
$,
\begin{equation}\label{eq:shift-and-emit}
 \cR_\chi(X)=V(X\otimes\chi)V^*\, ,
\end{equation}
which acts on tensor factors as
\[
\mathcal R_\chi(X_1\otimes\cdots\otimes X_n)
=V(X_1\otimes\cdots\otimes X_n\otimes\chi)V^*
=\underbrace{X_2\otimes\cdots\otimes X_n\otimes\chi}_{\text{new memory}}
\otimes\underbrace{X_1}_{\text{emitted system}}.
\]

To verify the output constraints, observe that after $k$
uses the retained memory consists of the last $n-k$
registers of the initial state, followed by $k$ copies
of $\chi$. Consequently, the marginal channels of
$\cR_\chi$ satisfy
\begin{equation}\label{eq:shift-marginals}
 \begin{aligned}
 \mathcal T_\chi^k(X)
   &=\tr_{A_1\cdots A_k}X\otimes\chi^{\otimes k},
       &&0\le k\le n,\\
 \mathcal O_\chi(\mathcal T_\chi^k(X))
   &=X_{A_{k+1}},
       &&0\le k<n.
 \end{aligned}
\end{equation}
Choosing the initial records $\omega_0=R_n$ and $\omega_1=T_n$, the quadruple $(M,\omega_0,\omega_1,\mathcal R_\chi)$ satisfies
\[
\mathcal O_\chi(\mathcal T_\chi^k(\omega_i))
=(\omega_i)_{A_{k+1}}=\rho_i,
\qquad i=0,1,\quad 0\le k<n\, ,
\]
so it defines an admissible $n$-use preparation with information cost $\mathbb D(R_n\Vert T_n)$. This establishes \(I_n^{\mathbb D}\le E_n\).

Finally, assume that $\mathbb D$ is jointly lower
semicontinuous and that its classical restriction is finite
for all pairs of probability distributions $p,q$ satisfying
$\supp p\subseteq\supp q$.
The extension sets $\cE_n(\rho_0)$ and
$\cE_n(\rho_1)$ are compact and nonempty.
To establish finiteness, recall that the support inclusion
$\supp\rho_0\subseteq\supp\rho_1$ guarantees the existence
of probability distributions $p,q$ with
$\supp p\subseteq\supp q$ and states $\tau_z$ such that
\[
 \rho_0=\sum_z p_z\tau_z,
 \qquad
 \rho_1=\sum_z q_z\tau_z.
\]
Consider the compatible extensions
\[
 R_n=\sum_z p_z\tau_z^{\otimes n},
 \qquad
 T_n=\sum_z q_z\tau_z^{\otimes n}.
\]
Since these states are obtained from the classical
distributions $p,q$ through a common channel, data
processing gives
\[
 \mathbb D(R_n\Vert T_n)
 \le \mathbb D(\operatorname{diag}p\Vert\operatorname{diag}q)
 <\infty.
\]
By joint lower semicontinuity of $\mathbb D$
and compactness of the sets of compatible extensions,
the infimum in \eqref{eq:horizon-extension-equivalence}
is therefore finite and attained by some pair
$(R_n^*,T_n^*)$.
This pair also attains the minimum
by choosing $M=S^{\otimes n}$ and $\Psi=\id$.
In both cases, a memory of dimension $(\dim S)^n$
suffices.
\end{proof}

\begin{theorem}[Interpolation and convergence of $n$-use preparation costs]
\label{AP-prop:hierarchy}\label{AP-thm:hierarchy}
Let $\mathbb D$ be a quantum divergence and let
$\rho_0,\rho_1\in\D(S)$ with $\supp\rho_0\subseteq\supp\rho_1$.
For every $n\ge1$,
\begin{equation}\label{AP-eq:horizon-general-hierarchy}
 \mathbb D(\rho_0\Vert\rho_1)=I_1^{\mathbb D}(\rho_0\Vert\rho_1) \le
 I_n^{\mathbb D}(\rho_0\Vert\rho_1)
   \le I_{n+1}^{\mathbb D}(\rho_0\Vert\rho_1)
    \le I_\infty^{\mathbb D}(\rho_0\Vert\rho_1)
    =\widehat{\mathbb D}(\rho_0\Vert\rho_1).
\end{equation}
If, in addition, $\widehat{\mathbb D}$ is jointly lower
semicontinuous at $(\rho_0,\rho_1)$ for pairs of states supported on
$\supp\rho_1$, then
\begin{equation}\label{AP-eq:hierarchy}
 \lim_{n\to\infty}I_n^{\mathbb D}(\rho_0\Vert\rho_1)
   =I_\infty^{\mathbb D}(\rho_0\Vert\rho_1)
   =\widehat{\mathbb D}(\rho_0\Vert\rho_1).
\end{equation}
\end{theorem}

\begin{proof}

The cases $n=1$ and $n=\infty$ are stated in \eqref{eq:horizon-endpoints}. The inequality  $I_n^{\mathbb D}(\rho_0\Vert\rho_1)
   \le I_{n+1}^{\mathbb D}(\rho_0\Vert\rho_1)$ follows from the fact that every preparation admissible for
$n+1$ uses is also admissible for $n$ uses, and every
indefinite-use preparation is admissible for each finite number of uses.
Taking infima yields the inequalities in
\eqref{AP-eq:horizon-general-hierarchy}. 

We must show now \eqref{AP-eq:hierarchy}. Suppose now that $\widehat{\mathbb D}$ is jointly lower semicontinuous at the target pair. Put $P=\supp\rho_1$ and $d=\operatorname{rank}\rho_1$. Since $\supp\rho_0\subseteq P$, both states $\rho_0$ and $\rho_1$ are supported on $PS$. As in the proof of Lemma~\ref{lemma:SupportRestriction}, let $\mathcal J$ denote the isometric inclusion of $PS$ into $S$, and let $\mathcal Q$ be a channel left inverse of $\mathcal J$. Composing the output of any preparation channel with $\mathcal Q$, or conversely with $\mathcal J$, preserves the memory dynamics and the information cost while restricting or embedding the prescribed output states. Hence all finite- and indefinite-use preparation costs are unchanged when $S$ is replaced by $PS$. We may therefore work on $PS$, where $\rho_1$ is faithful.

Fix an arbitrary admissible $n$-use preparation
$(M,\omega_0,\omega_1,\cR)$.
As in Section~\ref{subsec:parallel-preparation}, let
$
 \Psi_n=\tr_M\circ\cR^{[n]}:
 M\longrightarrow S^{\otimes n}
$
be the channel collecting the first $n$ emitted systems.
By \eqref{eq:sequential-output-marginals}, its individual
output marginals satisfy
\[
 [\Psi_n(X)]_{S_j}
 =\mathcal O(\mathcal T^{j-1}(X)),
 \qquad 1\le j\le n.
\]
Consequently, using the average-marginal channel
$\mathcal N_n$ introduced in
\eqref{eq:average-marginal}, we obtain
\[
 \mathcal N_n(\Psi_n(X))
 =\frac1n\sum_{j=1}^n[\Psi_n(X)]_{S_j}
 =\frac1n\sum_{k=0}^{n-1}
       \mathcal O(\mathcal T^k(X)).
\]
In particular, since the preparation is admissible
for $n$ uses, $
 \mathcal N_n(\Psi_n(\omega_i))=\rho_i
$. We now apply Lemma~\ref{lem:extraction} to the
collected output. The lemma provides a
measure-and-prepare channel
\[
 \Phi_n(Y)
 =\sum_z\tr(F_z^{(n)}Y)\tau_z^{(n)},
 \qquad Y\in\mathcal B(S^{\otimes n}),
\]
where $\{F_z^{(n)}\}_z$ is a POVM on
$S^{\otimes n}$ and $\tau_z^{(n)}\in\D(S)$,
satisfying
\[
 \|\Phi_n(Y)-\mathcal N_n(Y)\|_1
 \le\frac{2d^2}{n},
 \qquad Y\in\D(S^{\otimes n}).
\]

Composing $\Phi_n$ with $\Psi_n$ gives a channel
from the initial memory to the output system, $
 \mathcal C_{n,\cR}:=\Phi_n\circ\Psi_n.$ This channel is also measure and prepare. Indeed,
writing $\Psi_n^*$ for the adjoint of $\Psi_n$,
we have
\[
 \mathcal C_{n,\cR}(X)
 =\sum_z\tr(F_z^{(n)}\Psi_n(X))
       \tau_z^{(n)}
 =\sum_z\tr(E_z^{(n)}X)\tau_z^{(n)}\, ,
\]
where
$
 E_z^{(n)}:=\Psi_n^*(F_z^{(n)})
$. Since $\Psi_n$ is completely positive and trace
preserving, the operators $E_z^{(n)}$ are positive
and satisfy $\sum_zE_z^{(n)}=I_M$.
They therefore constitute a POVM on the
initial memory. Moreover, for every $X\in\D(M)$, the estimate
from Lemma~\ref{lem:extraction} gives
\[
 \|\mathcal C_{n,\cR}(X)
       -\mathcal N_n(\Psi_n(X))\|_1=
   \|\Phi_n(\Psi_n(X))
       -\mathcal N_n(\Psi_n(X))\|_1
 \le\frac{2d^2}{n}\, .
\]
Notice that the channel $\mathcal C_{n,\cR}$
depends on the preparation device through
$\Psi_n$, but the approximation error depends
only on $d$ and $n$.

Define the probability distributions
\[
 p_n(z)=\tr(E_z^{(n)}\omega_0),
 \qquad
 q_n(z)=\tr(E_z^{(n)}\omega_1),
\]
and the corresponding output states
\[
 (\rho_0')_n
   =\mathcal C_{n,\cR}(\omega_0)
     =\sum_zp_n(z)\tau_z^{(n)}\, , \quad
 (\rho_1')_n
   =\mathcal C_{n,\cR}(\omega_1)
     =\sum_zq_n(z)\tau_z^{(n)}\, .
\]
Applying the preceding estimate to
$\omega_0$ and $\omega_1$, respectively,
and recalling the prescribed output
conditions, we obtain
\begin{equation}\label{eq:general-horizon-approx}
 \|(\rho_0')_n-\rho_0\|_1
 +\|(\rho_1')_n-\rho_1\|_1
 \le\frac{4d^2}{n}.
\end{equation}

The distributions $p_n,q_n$ and states
$\{\tau_z^{(n)}\}_z$ thus define a common
classical reverse preparation of the
approximate pair $((\rho_0')_n,(\rho_1')_n)$.
Therefore,
\begin{equation}\label{eq:general-horizon-lower}
 \mathbb D(\omega_0\Vert\omega_1)
 \ge\mathbb D_{\mathrm{cl}}(p_n\Vert q_n)
 \ge\widehat{\mathbb D}((\rho_0')_n\Vert(\rho_1')_n)\, ,
\end{equation}
where the first inequality follows from data processing
under the measurement channel, and the second from the
variational characterization of the maximal extension
in \eqref{eq:maximal-extension}, since
$(p_n,q_n,\{\tau_z^{(n)}\}_z)$ is an admissible
classical reverse preparation of $((\rho_0')_n,(\rho_1')_n)$.

For $\delta\ge0$, define
\[
 m(\delta):=
 \inf_{\substack{a,b\in\D(PS)\\
   \|a-\rho_0\|_1+\|b-\rho_1\|_1\le\delta}}
   \widehat{\mathbb D}(a\Vert b).
\]
By \eqref{eq:general-horizon-approx}, the
pair $((\rho_0')_n,(\rho_1')_n)$ belongs to the
set defining $m(4d^2/n)$. Hence
\eqref{eq:general-horizon-lower} implies
\[
 \mathbb D(\omega_0\Vert\omega_1)
 \ge m(4d^2/n).
\]
Importantly, this lower bound holds for
every admissible $n$-use preparation,
regardless of its memory dimension.
Taking the infimum over all such
preparations therefore gives
\[
 m(4d^2/n)
 \le I_n^{\mathbb D}(\rho_0\Vert\rho_1)
 \le\widehat{\mathbb D}(\rho_0\Vert\rho_1).
\]

Finally, notice first that at \(\delta=0\), the only admissible pair is \((\rho_0,\rho_1)\), so
$
m(0)=\widehat{\mathbb D}(\rho_0\Vert\rho_1)\, .
$
 Joint lower semicontinuity of
$\widehat{\mathbb D}$ at $(\rho_0,\rho_1)$
implies
\[
 \lim_{\delta\downarrow0}m(\delta)
 =\widehat{\mathbb D}(\rho_0\Vert\rho_1).
\]
Since $4d^2/n\to0$, the preceding bounds
give
\[
 \lim_{n\to\infty}
 I_n^{\mathbb D}(\rho_0\Vert\rho_1)
 =\widehat{\mathbb D}(\rho_0\Vert\rho_1),
\]
as claimed.

\end{proof}

\begin{proposition}[Strict interpolation from recoverability]
\label{AP-prop:strict-hierarchy-general}
Let $\mathbb D$ be a quantum divergence, and let
$\rho_0,\rho_1\in\D(S)$ satisfy
$\supp\rho_0\subseteq\supp\rho_1$ and
$\widehat{\mathbb D}(\rho_0\Vert\rho_1)<\infty$.
Assume that the infimum defining $I_k^{\mathbb D}(\rho_0\Vert\rho_1)$
is attained for every finite $k$, and that $\mathbb D$ has the
following recovery property: whenever a channel $\Lambda$ and states
$a_0,a_1$ satisfy
\[
 \mathbb D(\Lambda(a_0)\Vert\Lambda(a_1))
 =\mathbb D(a_0 \Vert a_1)<\infty,
\]
there exists a recovery channel $\mathcal Q$ with
$\mathcal Q \circ \Lambda(a_0)=a_0$ and $\mathcal Q\circ \Lambda(a_1)=a_1$.
Then, for any distinct $n,m\in\mathbb N\cup\{+\infty\}$,
\begin{equation}\label{AP-eq:strict-general-hierarchy}
 I_n^{\mathbb D}(\rho_0\Vert\rho_1)
 =I_m^{\mathbb D}(\rho_0\Vert\rho_1)
 \quad\Longleftrightarrow\quad [\rho_0,\rho_1]=0.
\end{equation}
\end{proposition}

\begin{proof}
If $\rho_0$ and $\rho_1$ commute, their common eigenbasis gives a
classical reverse preparation with cost
$\mathbb D(\rho_0\Vert\rho_1)$. The classical restriction of
$\mathbb D$ and the definition of its maximal extension therefore
imply $\widehat{\mathbb D}(\rho_0\Vert\rho_1)
=\mathbb D(\rho_0\Vert\rho_1)$. The implication follows then from \eqref{eq:horizon-general-hierarchy}.

Conversely, suppose two distinct terms coincide. By  the monotonicity \eqref{eq:horizon-general-hierarchy},
there is a finite $n$ such that
$I_n^{\mathbb D}(\rho_0\Vert \rho_1)=I_{n+1}^{\mathbb D}(\rho_0\Vert \rho_1)=:c$, with the target pair
understood throughout. Choose an optimal $(n+1)$-use preparation
$(M,\omega_0,\omega_1,\cR)$, and write
$\mathcal T=\tr_S\circ\cR$ and $\mathcal O=\tr_M\circ\cR$.
The shifted memories $\eta_i=\mathcal T(\omega_i)$ are admissible
for $n$ uses, i.e. $
(M,\eta_0,\eta_1,\mathcal R)$ is an admissible quadruple for $n$ uses,
so data processing gives
\[
 c=I_n^{\mathbb D}(\rho_0\Vert \rho_1)
 \le\mathbb D(\eta_0\Vert\eta_1)
 \le\mathbb D(\omega_0\Vert\omega_1)=c.
\]
The assumed recovery property provides a channel $\mathcal Q$ with
$\mathcal Q(\eta_i)=\omega_i$. Hence
$\widetilde{\cR}=(\mathcal Q\otimes\id_S)\circ\cR$
preserves both initial memories, while its emitted marginals remain
$\rho_0$ and $\rho_1$.
Therefore  $(M,\omega_0,\omega_1,\widetilde{\cR})$ is a stationary preparation with cost $c$.
Theorem~\ref{thm:preparation} and the upper bound in
\eqref{eq:horizon-general-hierarchy} imply
$c=\widehat{\mathbb D}(\rho_0\Vert\rho_1)$.

Applying the classical reduction in the proof of
Theorem~\ref{thm:preparation} to this stationary preparation,
we obtain distributions $p,q$ and states $\tau_z$ with
$\rho_0=\sum_zp_z\tau_z$, $\rho_1=\sum_zq_z\tau_z$, and
$\mathbb D_{\mathrm{cl}}(p\Vert q)\le c$.
The opposite inequality follows from the reverse-preparation
characterization \eqref{eq:maximal-extension}; thus
$\mathbb D_{\mathrm{cl}}(p\Vert q)=c$.

Discard any labels with $p_z+q_z=0$ before defining the classical
register below. This preserves the reverse preparation and its divergence,
since adjoining and removing unused labels are quantum channels.

Consider the common classical-to-quantum channel
$\Gamma_n(\diag x)=\sum_zx_z\tau_z^{\otimes n}$,
and put $R_n=\Gamma_n(\diag p)$ and
$T_n=\Gamma_n(\diag q)$. These are compatible extensions of
$\rho_0$ and $\rho_1$, since $
(R_n)_{S_j}=\rho_0$, $
(T_n)_{S_j}=\rho_1$, for every  $j=1,\ldots,n
$. Proposition~\ref{prop:parallel-preparation}
and data processing therefore yield
\[
 c=I_n^{\mathbb D}(\rho_0 \Vert \rho_1)
 \le\mathbb D(R_n\Vert T_n)
 \le\mathbb D_{\mathrm{cl}}(p\Vert q)=c.
\]
Equality in data processing for $\Gamma_n$ implies that there exists a channel $\widehat{\mathcal Q}$ satisfying
$
\widehat{\mathcal Q}(R_n)=\operatorname{diag}p$, $
\widehat{\mathcal Q} (T_n)=\operatorname{diag}q
$.

Since the recovered states are diagonal, composing
$\widehat{\mathcal Q}$ with dephasing in the classical basis \[
\Delta(X)=\sum_y
|y\rangle\langle y|X|y\rangle\langle y|.
\]
preserves the recovery identities. By the
correspondence between quantum-to-classical
channels and POVMs \cite[Theorem~2.37]{Watrous2018},
the resulting channel is the measurement channel
associated with the effects
\begin{equation}\label{eq:Dual-Q-Hat}
E_y=\widehat{\mathcal Q}^*(|y\rangle\langle y|),
\qquad
\sum_y E_y=I,
\end{equation}
and acts as
\[
(\Delta\circ\widehat{\mathcal Q})(X)
=\sum_y\operatorname{tr}(E_yX)
|y\rangle\langle y|.
\]

Let $t_{y|z}=\tr(E_y\tau_z^{\otimes n})$. Since \(\{E_y\}\) is a POVM, the matrix \(t=(t_{y|z})\) is column-stochastic. Using that $\widehat{\mathcal Q}(R_n)=\diag p$, equation \eqref{eq:Dual-Q-Hat} and the  recoverability property, we obtain
\[
\begin{aligned}
p_y
&=\langle y, \widehat{\mathcal Q}(R_n)y\rangle\\
&=\operatorname{Tr}(E_yR_n)\\
&=\operatorname{Tr}\left(
E_y\sum_z p_z\tau_z^{\otimes n}\right)\\
&=\sum_z p_z\operatorname{Tr}
(E_y\tau_z^{\otimes n})\\
&=\sum_z t_{y|z}p_z\, .
\end{aligned}
\]
The resulting
column-stochastic matrix $t$ satisfies then $tp=p$. In a similar way we also show $tq=q$.

Set $r_z=(p_z+q_z)/2>0$ and let $\ell_z=p_z/r_z$ be the likelihood ratio of $p$ relative to $r$.
Since $tr=r$ and $tp=p$, we obtain
\[
\begin{aligned}
\sum_{y,z}r_z t_{y|z}(\ell_z-\ell_y)^2
&=\sum_z r_z\ell_z^2\sum_y t_{y|z}
-2\sum_y\ell_y\sum_z t_{y|z}p_z
+\sum_y\ell_y^2\sum_z t_{y|z}r_z\\
&=\sum_z r_z\ell_z^2
-2\sum_y\ell_y p_y
+\sum_y\ell_y^2r_y\\
&=\sum_z r_z\ell_z^2
-\sum_y r_y\ell_y^2=0.
\end{aligned}
\]
Thus $t_{y|z}>0$ implies $\ell_y=\ell_z$ whenever
$r_z>0$.

For each distinct value $\lambda$ of $\ell_z$, define
\[
F_\lambda=\sum_{y:\ell_y=\lambda}E_y.
\]
Since $t_{y|z}>0$ implies $\ell_y=\ell_z$ and
$\sum_y t_{y|z}=1$, the POVM $\{F_\lambda\}_\lambda$
satisfies
\[
\tr(F_\lambda\tau_z^{\otimes n})
=\sum_{y:\ell_y=\lambda}t_{y|z}
=\begin{cases}
1,&\ell_z=\lambda,\\
0,&\ell_z\neq\lambda.
\end{cases}
\]
Thus states belonging to different likelihood-ratio
classes are perfectly distinguishable and therefore
have orthogonal supports. In particular, whenever
$\ell_z\neq\ell_w$,
\[
0=\tr(\tau_z^{\otimes n}\tau_w^{\otimes n})
=\bigl(\tr(\tau_z\tau_w)\bigr)^n.
\]
Since $\tau_z,\tau_w\geq0$, this implies
$\tau_z\tau_w=0$.

For each distinct value $\ell$ of the likelihood ratio, define
\[
A_\ell=\sum_{z:\ell_z=\ell}r_z\tau_z.
\]
The orthogonality established above implies
$A_\ell A_{\ell'}=0$ whenever $\ell\neq\ell'$.
Moreover, using $p_z=r_z\ell_z$ and
$q_z=r_z(2-\ell_z)$, we obtain
\[
\rho_0=\sum_\ell \ell A_\ell,
\qquad
\rho_1=\sum_\ell(2-\ell)A_\ell.
\]
Consequently,
\[
\rho_0\rho_1
=\sum_\ell \ell(2-\ell)A_\ell^2
=\rho_1\rho_0,
\]
which proves that $\rho_0$ and $\rho_1$ commute.
\end{proof}

\begin{proposition}[Failure of recoverability for maximal divergences]
\label{AP-cor:maximal-recoverability}
Let $\widehat{\mathbb D}$ be a lower semicontinuous
maximal divergence. For every pair of noncommuting
states $\rho_0,\rho_1 \in \mathcal{D}(S)$ with $\operatorname{supp}\rho_0
\subseteq\operatorname{supp}\rho_1$ and
$\widehat{\mathbb D}(\rho_0\Vert\rho_1)<\infty$,
there exist states $R_2,T_2 \in \mathcal{D}(S^{\otimes 2})$ with marginals
\[
\tr_{S_j}R_2=\rho_0,\qquad
\tr_{S_j}T_2=\rho_1,\qquad j=1,2.
\]
For $\Phi=\tr_{S_2}$, they satisfy
\[
\widehat{\mathbb D}(\Phi(R_2)\Vert\Phi(T_2))
=\widehat{\mathbb D}(R_2\Vert T_2)
=\widehat{\mathbb D}(\rho_0\Vert\rho_1)<\infty,
\]
but no quantum channel $\mathcal Q$ satisfies
\[
\mathcal Q\circ\Phi(R_2)=R_2,
\qquad
\mathcal Q\circ\Phi(T_2)=T_2.
\]
\end{proposition}

\begin{proof}
Choose noncommuting states $\rho_0,\rho_1$ with
$\widehat{\mathbb D}(\rho_0\Vert\rho_1)<\infty$.
Since $\widehat{\mathbb D}$ is already maximal,
\eqref{eq:horizon-general-hierarchy} gives
\[
\widehat{\mathbb D}(\rho_0\Vert\rho_1)
=I_1^{\widehat{\mathbb D}}(\rho_0\Vert\rho_1)
\leq I_2^{\widehat{\mathbb D}}(\rho_0\Vert\rho_1)
\leq\widehat{\mathbb D}(\rho_0\Vert\rho_1).
\]
Thus all inequalities are equalities. By
Proposition~\ref{prop:parallel-preparation},
the two-use cost admits the representation
\[
I_2^{\widehat{\mathbb D}}(\rho_0\Vert\rho_1)
=
\inf_{\substack{
R_2,T_2\in\D(S^{\otimes2})\\
\tr_{S_j}R_2=\rho_0,\;
\tr_{S_j}T_2=\rho_1,\ j=1,2
}}
\widehat{\mathbb D}(R_2\Vert T_2).
\]
The feasible set is compact, and lower
semicontinuity ensures that the infimum is
attained by some compatible extensions
$R_2,T_2$. In particular,
$
\widehat{\mathbb D}(R_2\Vert T_2)
=\widehat{\mathbb D}(\rho_0\Vert\rho_1)
$. 
Taking $\Phi=\tr_{S_2}$, the marginal
constraints give $\Phi(R_2)=\rho_0$ and
$\Phi(T_2)=\rho_1$, and therefore
\[
\widehat{\mathbb D}(\Phi(R_2)\Vert\Phi(T_2))
=\widehat{\mathbb D}(R_2\Vert T_2).
\]
Suppose that a recovery channel $\mathcal Q$
existed. Then
$
\mathcal Q(\rho_0)=R_2$ and 
$\mathcal Q(\rho_1)=T_2$.
Since both marginals of $R_2$ and $T_2$
coincide with their respective target states,
$\mathcal Q$ would broadcast $\rho_0$ and $\rho_1$.
By the no-broadcasting theorem
\cite{BarnumCavesFuchsJozsaSchumacher1996},
this would imply $[\rho_0,\rho_1]=0$, contrary
to our choice. Hence no such recovery
channel exists.
\end{proof}

\subsection{Universal testing of preparation outputs}

\begin{theorem}[Universal Stein exponent for $n$-use preparations]
\label{AP-lem:fixed-block}
Let $\rho_0,\rho_1>0$ be states on $S$.
For every $n\ge1$ and $0<\varepsilon<1$, the limit
\begin{equation}\label{eq:AP-rn}
r_n(\rho_0\Vert\rho_1)
:=\lim_{m\to\infty}-\frac1m\log\beta_{n,m}^{\varepsilon}
\end{equation}
exists and is independent of $\varepsilon$.
The sequence $(r_n)_{n\ge1}$ is nondecreasing and satisfies
\begin{equation}\label{eq:AP-rn-basic-bounds}
D(\rho_0\Vert\rho_1)
=r_1(\rho_0\Vert\rho_1)
\le r_n(\rho_0\Vert\rho_1)
\le I_n^D(\rho_0\Vert\rho_1)
\le\BS(\rho_0\Vert\rho_1).
\end{equation}
Moreover, for $d=\dim S$ and  for
\[
 \lambda=\min\{\lambda_{\min}(\rho_0),\lambda_{\min}(\rho_1)\}\, , \qquad L_{\rho_0,\rho_1}=\log\frac{2}{\lambda}
             +\frac{2\sqrt{2}}{\lambda^{5/2}}\, ,
 \qquad n_0=\left\lceil\frac{4d^2}{\lambda}\right\rceil.
\]
the lower bound improves to
\begin{equation}\label{AP-eq:rn-bounds}
 r_n(\rho_0\Vert\rho_1)
 \ge\max\left\{
 D(\rho_0\Vert\rho_1),
 \BS(\rho_0\Vert\rho_1)-\frac{4d^2L_{\rho_0,\rho_1}}{n}
 \right\}.
\end{equation}
for every $n\geq n_0$.
\end{theorem}

\begin{proof}

\noindent\textit{Extension setup.}
Fix $n\geq1$ and write
$\mathcal A=\cE_n(\rho_0)$ and
$\mathcal B=\cE_n(\rho_1)$, which are compact
convex subsets of $\D(H)$ with $H=S^{\otimes n}$.
By Lemma~\ref{lem:device-output-pairs}, every pair
$R\in\mathcal A$ and $T\in\mathcal B$ can be realized
as the joint output states of a common admissible
$n$-use preparation. Since the same device is
independently initialized in each of the $m$ rounds,
the corresponding hypotheses are $R^{\otimes m}$
and $T^{\otimes m}$. The receiver does not know
the underlying preparation and must therefore
choose a single test $0\leq Q_m\leq I$ whose
error guarantees hold uniformly over all
admissible output states. More precisely, the
worst-case errors can be expressed as
\[
\begin{aligned}
\alpha_{n,m}(Q_m)
&=\sup_{R\in\mathcal A}
\tr\left[(I-Q_m)R^{\otimes m}\right],\\
\beta_{n,m}(Q_m)
&=\sup_{T\in\mathcal B}
\tr\left[Q_mT^{\otimes m}\right].
\end{aligned}
\]
Consequently, the optimal universal type-II
error under the constraint
$\alpha_{n,m}(Q_m)\leq\varepsilon$ takes the form
\[
\beta_{n,m}^{\varepsilon}
=
\inf_{\substack{
0\leq Q_m\leq I\\
\sup_{R\in\mathcal A}
\tr[(I-Q_m)R^{\otimes m}]\leq\varepsilon
}}
\sup_{T\in\mathcal B}
\tr[Q_mT^{\otimes m}].
\]
Thus the original optimization over preparation
devices reduces to a composite quantum
hypothesis-testing problem between the sets
$\mathcal A$ and $\mathcal B$.

\noindent\textit{Convexification and Fekete's lemma.} We first establish existence of its exponent. The alternative family
$\{T^{\otimes m}:T\in\mathcal B\}$ is not closed under tensor products
with different factors. Introduce the enlarged sets
\begin{equation}\label{eq:product-hull}
\mathcal G_m=\co \{T^{\otimes m}:T\in\mathcal B\}\ ,\qquad \mathcal F_m
 =\co\{T_1\otimes\cdots\otimes T_m:T_j\in\mathcal B\},
\end{equation}
and let $\widetilde\beta_m^\varepsilon$ denote the optimal error with
the original null family and alternative $\mathcal F_m$. The sets
$\mathcal F_m$ are compact and convex, are stable under partial traces, tensor
products, and permutations of the $m$ blocks, and contain
$\tau_*^{\otimes m}$ for the faithful state
$\tau_*=\rho_1^{\otimes n}\in\mathcal B$. They therefore satisfy the
five assumptions of the composite generalized Stein theorem
\cite[Theorem~1]{LamiRegulaTakagi2026}. Convexifying the null family
does not change the error constraint, since it is linear in the state.

For each fixed $R\in\mathcal A$, the sequence
\[
a_\ell(R):=\inf_{W\in\mathcal F_\ell}
D(R^{\otimes\ell}\Vert W)
\]
is subadditive, since
$\mathcal F_k\otimes\mathcal F_\ell
\subseteq\mathcal F_{k+\ell}$ and the Umegaki
relative entropy is additive under tensor products.
Moreover, $a_\ell(R)$ is finite because
$\mathcal F_\ell$ contains the faithful state
$\tau_*^{\otimes\ell}$. Thus, the cited theorem,
together with Fekete's lemma, gives, for every
$0<\varepsilon<1$,
\[
\begin{aligned}
\lim_{m\to\infty}
-\frac1m\log\widetilde\beta_m^\varepsilon
&=\inf_{R\in\mathcal A}
D^\infty(R\Vert\mathcal F)\\
&=\inf_{R\in\mathcal A}
\lim_{\ell\to\infty}\frac1\ell
\inf_{W\in\mathcal F_\ell}
D(R^{\otimes\ell}\Vert W)\\
&=\inf_{R\in\mathcal A}
\inf_{\ell\ge1}\frac1\ell
\inf_{W\in\mathcal F_\ell}
D(R^{\otimes\ell}\Vert W)
=:\widetilde r_n.
\end{aligned}
\]

\noindent\textit{Symmetrization.} Since \(\mathcal B\) is compact and convex and contains the faithful state \(\tau_*=\rho_1^{\otimes n}\), the finite operator-order approximation in the proof of \cite[Prop.~7]{Lami2025} applies directly. Indeed, for every \(T\in\mathcal B\) and \(\delta>0\), the state
\[
s_T=e^{-\delta}T+(1-e^{-\delta})\tau_*
\in\mathcal B
\]
satisfies
\[
e^\delta s_T-T=(e^\delta-1)\tau_*>0.
\]
These inequalities define an open cover of \(\mathcal B\), from which compactness provides finitely many states \(s_1,\ldots,s_{N_\delta}\in\mathcal B\) such that every \(T\in\mathcal B\) satisfies \(T\le e^\delta s_j\) for some \(j\).

Following the remaining steps of the proof of
\cite[Prop.~7]{Lami2025},
we apply the preceding operator inequalities to
tensor products of states in $\mathcal B$.
For $T_1,\ldots,T_m\in\mathcal B$, choose
$s_{j_1},\ldots,s_{j_m}$ from the finite family
such that $T_k\leq e^\delta s_{j_k}$. Then
\[
T_1\otimes\cdots\otimes T_m
\leq e^{m\delta}
s_{j_1}\otimes\cdots\otimes s_{j_m}.
\]

Let $H=S^{\otimes n}$ and define the block-twirling
channel
\[
\mathcal T_m^{\mathrm{blk}}(X)
=\frac{1}{m!}\sum_{\pi\in\Sigma_m}U_\pi XU_\pi^*,
\qquad X\in\mathcal L(H^{\otimes m}),
\]
where $U_\pi$ permutes the $m$ blocks $H$.
Applying the operator domination estimate of
\cite[Eq.~(52)]{Lami2025} to the finite family
constructed above, we obtain, for every
$W\in\mathcal F_m$, a state $V\in\mathcal G_m$
such that
\begin{equation}\label{eq:type-domination}
\mathcal T_m^{\mathrm{blk}}(W)
\leq e^{m\delta}(m+1)^{N_\delta}V.
\end{equation}
The exponential factor comes from the domination
of each tensor factor, while the polynomial factor
follows from the method of types after permutation
averaging. Importantly, $N_\delta$ is independent
of $m$.

\noindent\textit{Existence of the limit.} The original i.i.d. families are permutation invariant, so tests
may be symmetrized without changing their errors. The inclusion
$\mathcal G_m\subseteq\mathcal F_m$ and
\eqref{eq:type-domination} imply
\begin{equation}\label{eq:beta-comparison}
 \beta_{n,m}^\varepsilon
 \le\widetilde\beta_m^\varepsilon
 \le e^{m\delta}(m+1)^{N_\delta}\beta_{n,m}^\varepsilon.
\end{equation}
Since the enlarged exponent equals $\widetilde r_n$, taking
logarithms and dividing by $m$ gives
\[
 \widetilde r_n
 \le\liminf_{m\to\infty}-\frac1m\log\beta_{n,m}^\varepsilon
 \le\limsup_{m\to\infty}-\frac1m\log\beta_{n,m}^\varepsilon
 \le\widetilde r_n+\delta.
\]
Here $N_\delta\log(m+1)/m\to0$ at fixed $n$ and $\delta$.
Letting $\delta\downarrow0$ proves existence and independence of
$\varepsilon$, as well as the characterization
\begin{equation}\label{eq:enlarged-limit}
 r_n(\rho_0\Vert\rho_1)
 =\inf_{R\in\mathcal A}\inf_{\ell\ge1}
   \frac1\ell\inf_{W\in\mathcal F_\ell}
   D(R^{\otimes\ell}\Vert W).
\end{equation}

\noindent\textit{The bounds $D\leq r_n\leq I_n$.}

Let \(\Lambda_n=\operatorname{tr}_{S_2\cdots S_n}\) denote the channel retaining only the first emission of each block. By the definition of the extension sets, \(\Lambda_n(R)=\rho_0\) for every \(R\in\mathcal A\), and \(\Lambda_n(T)=\rho_1\) for every \(T\in\mathcal B\). Moreover, each \(W\in\mathcal F_\ell\) can be expressed as a convex combination
\[
W=\sum_a p_a\,T_{a,1}\otimes\cdots\otimes T_{a,\ell},
\qquad T_{a,j}\in\mathcal B.
\]

Since \(\Lambda_n(T_{a,j})=\rho_1\) for all \(a,j\), applying \(\Lambda_n\) independently to the \(\ell\) blocks gives
\[
\begin{aligned}
\Lambda_n^{\otimes\ell}(R^{\otimes\ell})
&=\rho_0^{\otimes\ell},\\
\Lambda_n^{\otimes\ell}(W)
&=\sum_a p_a\bigotimes_{j=1}^{\ell}
\Lambda_n(T_{a,j})
=\rho_1^{\otimes\ell}.
\end{aligned}
\]
Data processing and additivity of Umegaki relative entropy therefore yield
\[
\begin{aligned}
D(R^{\otimes\ell}\Vert W)
&\ge D\bigl(\Lambda_n^{\otimes\ell}(R^{\otimes\ell})
\Vert\Lambda_n^{\otimes\ell}(W)\bigr)\\
&=D(\rho_0^{\otimes\ell}\Vert\rho_1^{\otimes\ell})\\
&=\ell D(\rho_0\Vert\rho_1).
\end{aligned}
\]
Since this inequality holds for every \(R\in\mathcal A\), \(W\in\mathcal F_\ell\), and \(\ell\ge1\), taking the infima in \(\eqref{eq:enlarged-limit}\) proves
\[
r_n(\rho_0\Vert\rho_1)\ge D(\rho_0\Vert\rho_1).
\]

Conversely, restricting the infimum in \(\eqref{eq:enlarged-limit}\) to \(\ell=1\) and observing that \(\mathcal F_1=\mathcal B\), because \(\mathcal B\) is convex, gives
\[
\begin{aligned}
r_n(\rho_0\Vert\rho_1)
&\le \inf_{R\in\mathcal A}\inf_{W\in\mathcal F_1}
D(R\Vert W)\\
&=\inf_{\substack{R\in\mathcal A\\T\in\mathcal B}}
D(R\Vert T)\\
&=I_n^D(\rho_0\Vert\rho_1),
\end{aligned}
\]
where the last equality follows from Proposition~\ref{prop:parallel-preparation}. 

In addition, since $D(\rho_0\Vert \rho_1)=I_1^D(\rho_0\Vert \rho_1)$, we obtain that $r_1(\rho_0\Vert \rho_1)=D(\rho_0\Vert \rho_1)$.

\noindent\textit{Monotonicity: $r_{n+1}\geq r_n$.}

Let \(\Lambda_n=\operatorname{Tr}_{S_{n+1}}\) denote the channel discarding the last emission of an \((n+1)\)-use block. The marginal constraints imply
\[
\Lambda_n(\mathcal E_{n+1}(\rho_i))
\subseteq\mathcal E_n(\rho_i),
\qquad i=0,1.
\]
Hence, given any feasible universal test \(Q_{n,m}\) for \(m\) rounds of \(n\) emissions, the test
\[
\widetilde Q_{n+1,m}
=(\Lambda_n^{\otimes m})^*(Q_{n,m})
\]
on the longer blocks satisfies
\[
\begin{aligned}
\alpha_{n+1,m}(\widetilde Q_{n+1,m})&=\sup_{X\in\mathcal E_{n+1}(\rho_0)}
\operatorname{tr}\left[
(I-Q_{n,m})\Lambda_n^{\otimes m}(X^{\otimes m})
\right]\\
&=\sup_{X\in\mathcal E_{n+1}(\rho_0)}
\operatorname{tr}\left[
(I-Q_{n,m})(\Lambda_n(X))^{\otimes m}
\right]\\
&\leq\sup_{R\in\mathcal E_n(\rho_0)}
\operatorname{tr}\left[
(I-Q_{n,m})R^{\otimes m}
\right]\\
&=\alpha_{n,m}(Q_{n,m}).
\end{aligned}
\]
Analogously,
\[
\beta_{n+1,m}(\widetilde Q_{n+1,m})
\leq\beta_{n,m}(Q_{n,m}).
\]
Taking the infimum over feasible tests gives \(\beta_{n+1,m}^\varepsilon\leq\beta_{n,m}^\varepsilon\). The monotonicity \(r_{n+1}\geq r_n\) then follows by taking the limit \(m\to\infty\).

\noindent\textit{The lower bound \eqref{AP-eq:rn-bounds}.}
Fix $n\geq n_0$ and let $\mathcal M_n$ be the
measurement channel associated with the POVM
$\{F_z^{(n)}\}_z$ of Lemma~\ref{lem:extraction}.
For $R\in\mathcal A$ and $T\in\mathcal B$,
denote the corresponding outcome distributions by
\[
p_R(z)=\tr(F_z^{(n)}R),\qquad
q_T(z)=\tr(F_z^{(n)}T).
\]
The proof of Lemma~\ref{lem:quantitative-horizon},
before minimizing over compatible extensions,
establishes the uniform estimate
\begin{equation}\label{eq:classical-uniform-lower}
\KL(p_R\Vert q_T)
\geq \BS(\rho_0\Vert\rho_1)
-\frac{4d^2L_{\rho_0,\rho_1}}{n}
=:h_n,
\qquad R\in\mathcal A,\ T\in\mathcal B.
\end{equation}
We now show that this lower bound also holds
per block for the regularized alternatives in
\eqref{eq:enlarged-limit}.

To this end, fix $R\in\mathcal A$,
$\ell\geq1$, and $W\in\mathcal F_\ell$.
By the definition of $\mathcal F_\ell$, we may write
\[
W=\sum_a\theta_a\,
T_{a,1}\otimes\cdots\otimes T_{a,\ell},
\qquad
T_{a,j}\in\mathcal B,
\]
where $\theta_a\geq0$ and $\sum_a\theta_a=1$.
Applying $\mathcal M_n$ independently to the
$\ell$ blocks transforms the null hypothesis into
$p_R^{\otimes\ell}$ and the alternative into
the classical distribution
\[
w=\sum_a\theta_a\,
q_{T_{a,1}}\otimes\cdots\otimes q_{T_{a,\ell}}\, ,  \qquad
w(z_1,\ldots,z_\ell)=
\sum_a\theta_a\prod_{j=1}^{\ell}q_{T_{a,j}}(z_j)\, .
\]
Although $w$ need not be a product distribution,
its conditional distributions remain in the
convex family
$\mathcal Q_n=\{q_T:T\in\mathcal B\}$.

Indeed, for 
$z_{<j}=(z_1,\ldots,z_{j-1})$
with $w(z_{<j})>0$, define
\[
\theta_a(z_{<j})
=
\frac{\theta_a\prod_{k<j}
q_{T_{a,k}}(z_k)}
{w(z_{<j})}.
\]
These coefficients are nonnegative and sum to one.
The conditional distribution of the $j$-th
outcome can therefore be expressed as
\[
w_j(\,\cdot\mid z_{<j})=\frac{w(z_{<j},\,\cdot)}{w(z_{<j})}=\frac{\sum_a\theta_a
\left(\prod_{k<j}q_{T_{a,k}}(z_k)\right)
q_{T_{a,j}}(\,\cdot\, )}
{w(z_{<j})}
=\sum_a\theta_a(z_{<j})q_{T_{a,j}}(\,\cdot\, )=q_{\overline T_j(z_{<j})}(\,\cdot\, ),
\]
where
\[
\overline T_j(z_{<j})
=\sum_a\theta_a(z_{<j})T_{a,j}
\in\mathcal B.
\]
The last inclusion follows from the convexity
of $\mathcal B$, while the second equality
uses the linearity of the measurement map.
Consequently, \eqref{eq:classical-uniform-lower}
applies to each conditional distribution and gives
\[
\KL\bigl(p_R\Vert
w_j(\,\cdot\mid z_{<j})\bigr)\geq h_n.
\]

We can now apply data processing under
$\mathcal M_n^{\otimes\ell}$, followed by the
classical relative-entropy chain rule, to obtain
\[
\begin{aligned}
D(R^{\otimes\ell}\Vert W)
&\geq\KL(p_R^{\otimes\ell}\Vert w)\\
&=\sum_{z_1,\ldots,z_\ell}
\left(\prod_{k=1}^{\ell}p_R(z_k)\right)
\log
\prod_{j=1}^{\ell}
\frac{p_R(z_j)}{w_j(z_j\mid z_{<j})}\\
&=\sum_{j=1}^{\ell}
\sum_{z_1,\ldots,z_\ell}
\left(\prod_{k=1}^{\ell}p_R(z_k)\right)
\log\frac{p_R(z_j)}{w_j(z_j\mid z_{<j})}\\
&=\sum_{j=1}^{\ell}
\sum_{z_{<j}}
p_R^{\otimes(j-1)}(z_{<j})
D_{\mathrm{KL}}\!\left(
p_R\middle\Vert w_j(\cdot\mid z_{<j})
\right)\\
&=\sum_{j=1}^{\ell}
\mathbb E_{Z_{<j}\sim p_R^{\otimes(j-1)}}
\KL\!\left(
p_R\middle\Vert
w_j(\,\cdot\mid Z_{<j})
\right)\\
&\geq\sum_{j=1}^{\ell}h_n
=\ell h_n.
\end{aligned}
\]

If the classical relative entropy is infinite,
the lower bound is immediate. Otherwise,
all conditional distributions appearing in the
expectation are well defined on the relevant
histories.

Since this estimate holds for every
$R\in\mathcal A$, $W\in\mathcal F_\ell$ and
$\ell\geq1$, taking the infima in
\eqref{eq:enlarged-limit} yields
\[
r_n(\rho_0\Vert\rho_1)
\geq\BS(\rho_0\Vert\rho_1)
-\frac{4d^2L_{\rho_0,\rho_1}}{n},
\qquad n\geq n_0.
\]
Together with \eqref{eq:rn-basic-bounds},
this proves \eqref{AP-eq:rn-bounds}.

\end{proof}

\begin{theorem}[Belavkin--Staszewski Stein limit]
\label{AP-thm:BS-Stein}
Let $\rho_0,\rho_1>0$ be states on $S$, and let
$0<\varepsilon<1$. The universal exponents of
Theorem~\ref{lem:fixed-block} satisfy
\begin{equation}\label{AP-eq:iterated-Stein}
 \lim_{n\to\infty}r_n(\rho_0\Vert\rho_1)
 =\BS(\rho_0\Vert\rho_1).
\end{equation}
More generally, for every sequence of positive integers
$n_m\to\infty$,
\begin{equation}\label{AP-eq:simultaneous-Stein}
 \lim_{m\to\infty}-\frac1m\log\beta_{n_m,m}^{\varepsilon}
 =\BS(\rho_0\Vert\rho_1).
\end{equation}
\end{theorem}

\begin{proof}
The first limit is an immediate consequence of
Theorem~\ref{lem:fixed-block}: for $n\ge n_0$, its bounds imply
\[
 0\le \BS(\rho_0\Vert\rho_1)-r_n(\rho_0\Vert\rho_1)
 \le \frac{4d^2L_{\rho_0,\rho_1}}{n}.
\]
It remains to establish the simultaneous limit without imposing a
relation between $n_m$ and $m$.

Fix \(k\ge1\). Since \(n_m\to\infty\), we have \(n_m\ge k\) for all sufficiently large \(m\). Every admissible \(n_m\)-use preparation is also admissible for \(k\) uses. Recall that
\[
\Omega_{i,n}^{\mathsf P}
=\operatorname{Tr}_M\mathcal R^{[n]}(\omega_i),
\]
where \(\mathcal R^{[n]}:M\to M\otimes S^{\otimes n}\) describes \(n\) successive applications of \(\mathcal R\), leaving the previously emitted systems untouched.
By construction,
$
\mathcal R^{[j+1]}
=
(\mathcal R\otimes\operatorname{id}_{S^{\otimes j}})
\circ\mathcal R^{[j]}
$, 
up to the ordering of the tensor factors. Since \(\mathcal R\) is trace preserving, for any operator \(X\) on \(M\otimes S^{\otimes j}\),
\[
\operatorname{Tr}_{M S_{j+1}}
\bigl[(\mathcal R\otimes\operatorname{id}_{S^{\otimes j}})(X)\bigr]
=
\operatorname{Tr}_M X.
\]
Applying this to \(X=\mathcal R^{[j]}(\omega_i)\), we obtain
\[
\begin{aligned}
\operatorname{Tr}_{S_{j+1}}
\Omega_{i,j+1}^{\mathsf P}
&=\operatorname{Tr}_{M S_{j+1}}
\mathcal R^{[j+1]}(\omega_i)\\
&=\operatorname{Tr}_M\mathcal R^{[j]}(\omega_i)\\
&=\Omega_{i,j}^{\mathsf P}.
\end{aligned}
\]
This identity holds for every \(j\ge1\). Applying it repeatedly, starting from \(j=n_m-1\) and ending at \(j=k\), gives
\[
\operatorname{Tr}_{S_{k+1}\cdots S_{n_m}}
\Omega_{i,n_m}^{\mathsf P}
=
\Omega_{i,k}^{\mathsf P}.
\]

Denote this partial-trace channel by \(\Lambda_{n_m,k}\). Given any feasible universal test \(0\le Q_{k,m}\le I\) for \(m\) rounds of \(k\) emissions, define a test on the longer blocks by
\[
\widetilde Q_{n_m,m}
=\bigl(\Lambda_{n_m,k}^{\otimes m}\bigr)^*(Q_{k,m}),
\]
where the adjoint simply extends the test by the identity on the discarded systems. Since \(\mathfrak P_{n_m}(\rho_0,\rho_1)\subseteq\mathfrak P_k(\rho_0,\rho_1)\), the corresponding worst-case errors satisfy
\[
\begin{aligned}
\alpha_{n_m,m}(\widetilde Q_{n_m,m})
&\le \alpha_{k,m}(Q_{k,m})\le\varepsilon,\\
\beta_{n_m,m}(\widetilde Q_{n_m,m})
&\le \beta_{k,m}(Q_{k,m}).
\end{aligned}
\]

Thus \(\widetilde Q_{n_m,m}\) is feasible for the longer experiment and has no larger worst-case type-II error. Taking the infimum over feasible \(Q_{k,m}\) yields
\[
\beta_{n_m,m}^{\varepsilon}
\le\beta_{k,m}^{\varepsilon}.
\]
Consequently, the fixed-\(k\) limit from Theorem~\ref{lem:fixed-block} gives
\[
\liminf_{m\to\infty}
-\frac1m\log\beta_{n_m,m}^{\varepsilon}
\ge
\lim_{m\to\infty}
-\frac1m\log\beta_{k,m}^{\varepsilon}
=r_k(\rho_0\Vert\rho_1).
\]
As $k$ is arbitrary, the first limit shows that the left-hand
side is at least $\BS(\rho_0\Vert\rho_1)$.

For the converse, consider the optimal classical reverse preparation
$p,q,\{\tau_z\}$ satisfying 
$\rho_0=\sum_zp_z\tau_z$, $\rho_1=\sum_zq_z\tau_z$, and
$\KL(p\Vert q)=\BS(\rho_0\Vert\rho_1)$.
The stationary channel \eqref{eq:classical-stationary-device}
with initial memories $\diag p$ and $\diag q$ is admissible for
every number $n$ of emissions. Its output states are
\[
 R_n=\sum_zp_z\tau_z^{\otimes n},\qquad
 T_n=\sum_zq_z\tau_z^{\otimes n},
\]
which are obtained from the classical distributions by the same
channel $\Gamma_n(\diag x)=\sum_zx_z\tau_z^{\otimes n}$.

Let $Q_m$ be any feasible universal  test on $m$
independent $n$-use rounds. Define the pulled-back effect
\[
\widetilde Q_m=(\Gamma_n^{\otimes m})^*(Q_m).
\]
Since $\Gamma_n$ is a classical-to-quantum channel,
its adjoint maps effects to classical effects. Hence
$0\leq\widetilde Q_m\leq I$ defines a test on the
classical register. By construction,
\[
\begin{aligned}
&\operatorname{tr}\left[
(I-\widetilde Q_m)(\operatorname{diag}p)^{\otimes m}
\right]
=\operatorname{tr}\left[
(I-Q_m)R_n^{\otimes m}
\right],\\
&\operatorname{tr}\left[
\widetilde Q_m(\operatorname{diag}q)^{\otimes m}
\right]
=\operatorname{tr}\left[
Q_mT_n^{\otimes m}
\right].
\end{aligned}
\]
Since the chosen stationary classical preparation is
admissible for every $n$, these quantities are bounded
above by $\alpha_{n,m}(Q_m)\leq\varepsilon$ and
$\beta_{n,m}(Q_m)$, respectively. Thus
$\widetilde Q_m$ is a feasible classical test, and
optimality of the classical type-II error gives
\[
\beta_{\mathrm{cl},m}^{\varepsilon}(p,q)
\leq
\operatorname{tr}\left[
\widetilde Q_m(\operatorname{diag}q)^{\otimes m}
\right]
\leq\beta_{n,m}(Q_m).
\]
Taking the infimum over feasible universal tests yields
\[
\beta_{n,m}^{\varepsilon}
\geq\beta_{\mathrm{cl},m}^{\varepsilon}(p,q).
\]
Since this comparison holds for every $n,m$, it remains
valid along any sequence $n=n_m$. The classical Stein
lemma therefore gives
\[
\limsup_{m\to\infty}
-\frac{1}{m}\log\beta_{n_m,m}^{\varepsilon}
\leq
\lim_{m\to\infty}
-\frac{1}{m}
\log\beta_{\mathrm{cl},m}^{\varepsilon}(p,q)=D_{\mathrm{KL}}(p\Vert q)
=\BS(\rho_0\Vert\rho_1),
\]
which proves the required converse bound.

\end{proof}

\section{Auxiliary results}

\subsection{Reduction to the joint memory support}

\begin{lemma}[Reduction to the joint memory support]
\label{lemma:SupportRestriction}
Let $\cR:\cL(M)\to\cL(M\otimes S)$ be a quantum channel, and let
$\omega_0,\omega_1\in\D(M)$ satisfy
$\mathcal T(\omega_i)=\omega_i$ for $i=0,1$, where
$\mathcal T=\tr_S\circ\cR$.
Let $P$ be the orthogonal projection onto $\supp(\omega_0+\omega_1)$, $M_P=PM$, and let
$\iota:M_P\hookrightarrow M$ be the natural inclusion. There is a
channel $\widetilde{\cR}:\cL(M_P)\to\cL(M_P\otimes S)$ such that,
for $\widetilde\omega_i=\iota^*\omega_i\iota$,
\[
 \tr_S\widetilde{\cR}(\widetilde\omega_i)
   =\widetilde\omega_i,\qquad
 \tr_{M_P}\widetilde{\cR}(\widetilde\omega_i)
   =\tr_M\cR(\omega_i),\qquad i=0,1.
\]
Moreover, for every quantum divergence $\mathbb D$ satisfying data
processing,
\[
 \mathbb D(\widetilde\omega_0\|\widetilde\omega_1)
 =\mathbb D(\omega_0\|\omega_1).
\]
In particular, the memory may be restricted to the joint support of
the two states.
\end{lemma}

\begin{proof}
Let
$P$ be the orthogonal projection onto $\supp(\omega_0+\omega_1)$, write $M_P=PM$, and put
$\overline\omega=(\omega_0+\omega_1)/2$. Choose a Stinespring isometry
$V:M\to M\otimes S\otimes E$ for $\mathcal R$,
where $E$ is an auxiliary environment, so that
\[
\mathcal R(X)=\tr_E(VXV^*),
\qquad
\mathcal T(X)=\tr_{S\otimes E}(VXV^*).
\] Then
$\omega_i=P\omega_iP$, the state $\overline\omega$ is faithful on
$M_P$, and $\mathcal T(\overline\omega)=\overline\omega$. With
$\Pi_\perp=(I_M-P)\otimes I_{S\otimes E}$, we have
 \begin{equation}
  0 =\tr[(I_M-P)\overline\omega]
     =\tr[\Pi_\perp V\overline\omega V^*]=\bigl\|\Pi_\perp V\overline\omega^{1/2}\bigr\|_2^2.
 \end{equation}
As $\overline\omega^{1/2}$ has range $M_P$, this implies
$\Pi_\perp VP=0$. Hence $V$ maps $M_P$ isometrically into
$M_P\otimes S\otimes E$, and its restriction defines a channel on
$M_P$ which still preserves both restricted memory states.

We may therefore restrict the memory Hilbert space to \(M_P=PM\), since both memory states are supported on this subspace and the preceding argument shows that the retained memory cannot leave it. It remains to verify that this restriction does not change their information cost. Let \(\iota:M_P\to M\) denote the natural isometric inclusion, satisfying \(\iota^*\iota=I_{M_P}\) and \(\iota\iota^*=P\), and write \(\widetilde\omega_i=\iota^*\omega_i\iota\).
The channel \(\mathcal J(X)=\iota X\iota^*\) embeds states on \(M_P\) into \(M\), with \(\mathcal J(\widetilde\omega_i)=\omega_i\). Conversely, for any fixed state \(\tau\in\mathcal D(M_P)\), define
\[
\mathcal Q(X)=\iota^*X\iota+\operatorname{tr}[(I_M-P)X]\,\tau.
\]
This map is completely positive and trace preserving: it compresses the input onto \(M_P\) and replaces the weight on its orthogonal complement by \(\tau\). Moreover, \(\mathcal Q\circ\mathcal J=\operatorname{id}_{M_P}\).

Thus the original memory states and their restrictions can be transformed into one another by quantum channels. Applying data processing first to \(\mathcal J\) and then to \(\mathcal Q\), we obtain
\[
\begin{aligned}
\mathbb D(\omega_0\Vert\omega_1)
&=\mathbb D(\mathcal J(\widetilde\omega_0)\Vert
\mathcal J(\widetilde\omega_1))\\
&\le\mathbb D(\widetilde\omega_0\Vert\widetilde\omega_1)\\
&=\mathbb D(\mathcal Q(\omega_0)\Vert
\mathcal Q(\omega_1))\\
&\le\mathbb D(\omega_0\Vert\omega_1).
\end{aligned}
\]
Hence \(\mathbb D(\widetilde\omega_0\Vert\widetilde\omega_1)=\mathbb D(\omega_0\Vert\omega_1)\). 
\end{proof}

\subsection{Uniform extraction from correlated emissions}\label{app:uniform-extraction}

Introduce the permutation-twirling channel and its average
one-site marginal:
\begin{align}
 \mathcal T_n(X)
   &=\frac1{n!}\sum_{\pi\in\Sigma_n}U_\pi XU_\pi^*,
       \label{eq:twirl}\\
 \mathcal N_n(X)
   &=\tr_{S_2\cdots S_n}\mathcal T_n(X)
     =\frac1n\sum_{j=1}^n\tr_{S_{\ne j}}X.
       \label{eq:average-marginal}
\end{align}

\begin{lemma}[Uniform extraction]\label{lem:extraction}
For each $n$ there are a finite POVM $\{F_z^{(n)}\}_{z\in Z_n}$ on
$S^{\otimes n}$ and states $\tau_z^{(n)}\in\D(S)$ such that the channel
\begin{equation}\label{eq:Phi-n}
 \Phi_n(X)=\sum_{z\in Z_n}\tr(F_z^{(n)}X)\tau_z^{(n)}
\end{equation}
satisfies
\begin{equation}\label{eq:channel-approx}
 \norm{\Phi_n(X)-\mathcal N_n(X)}_1\le\frac{2d^2}{n}
 \quad\text{for every }X\in\D(S^{\otimes n}),
 \qquad d=\dim S.
\end{equation}
\end{lemma}
\begin{proof}
To show this result, we will make use of \cite[Theorem~5]{Chiribella2011}. We will verify first that the hypothesis are fulfilled. Firstly, the twirling channel $\mathcal{T}_n$ is permutation invariant, i.e. $U_{\sigma}\mathcal{T}_n(X)U_{\sigma}^*=\mathcal{T}_n(X)$ for every $\sigma \in \Sigma_n$. Moreover, its first output marginal is precisely $\mathcal{N}_n$. Indeed, each tensor factor appears in the $j$-th position $(n-1)!$ times. Consequently,
\begin{equation}
     \mathcal N_n(X)
   =\tr_{S_2\cdots S_n}\mathcal T_n(X)=\frac{1}{n!}\sum_{\sigma \in \Sigma_n}\tr_{S_2\cdots S_n}[U_{\sigma}X U_{\sigma}^*]
     =\frac1n\sum_{j=1}^n\tr_{S_{\ne j}}X
\end{equation}

Applying now  \cite[Theorem~5]{Chiribella2011}, there exists an entanglement breaking channel $\widetilde{\mathcal{T}_n}:\mathcal{B}(S^{\otimes n})\to \mathcal{B}(S^{\otimes n})$ such that 
\begin{equation}
    \Vert \tr_{S_2\cdots S_n} \widetilde{\mathcal{T}_n}-\mathcal{N}_n\Vert_{\diamond}\leq \frac{2d^2}{n}\, .
\end{equation}
Define $\Phi_n=\tr_{S_2\cdots S_n} \widetilde{\mathcal{T}_n}$. Since $\widetilde{\mathcal{T}_n}$ is entanglement breaking, so is $\Phi_n$. Using the representation of the entanglement breaking channels \eqref{eq:MP}, we can write
\[
\Phi_n(X)=\sum_{z^{(n)}} \tr[F_z^{(n)}X]\tau_z^{(n)}\, ,
\]
where $\{F_z^{(n)}\}$ is a POVM and $\tau_z^{(n)}$ are states on $S$. Finally, for every $X\in \mathcal{D}(S^{\otimes n})$, the diamond-norm estimate gives
\[
\Vert \Phi_n(X)-\mathcal{N}_n(X)\Vert_1\leq \Vert \Phi_n-\mathcal{N}_n\Vert_{\diamond}\leq\frac{2d^2}{n}\, .
\]
\end{proof}

\begin{lemma}[Quantitative bounds at finite $n$]
\label{lem:quantitative-horizon}
Let $\rho,\sigma>0$ on $S$, with $d=\dim S$, and let
$\mathbb D$ satisfy data processing and have classical restriction
$\mathbb D_{\mathrm{cl}}=D_{\mathrm{KL}}$. Set
\[
 \lambda=\min\{\lambda_{\min}(\rho),\lambda_{\min}(\sigma)\},
 \qquad n_0=\left\lceil\frac{4d^2}{\lambda}\right\rceil,
\]
and choose the explicit constant
\[
 L_{\rho,\sigma}=\log\frac{2}{\lambda}
 +\frac{2\sqrt{2}}{\lambda^{5/2}}\, .
\]
Then
\begin{equation}\label{eq:Bn-rate}
 0\le\BS(\rho\Vert\sigma)-I_n^{\mathbb D}(\rho\Vert\sigma)
 \le\frac{4d^2L_{\rho,\sigma}}{n},\qquad n\ge n_0.
\end{equation}
\end{lemma}

\begin{proof}
Let $\Phi_n$ and $\mathcal N_n$ be the channels of
Lemma~\ref{lem:extraction}. For compatible extensions $R_n,T_n$,
the marginal constraints give
$\mathcal N_n(R_n)=\rho$ and $\mathcal N_n(T_n)=\sigma$.
Writing $\Phi_n(R_n)$ and $\Phi_n(T_n)$ in terms of a common joint preparation,
\[
 a_n=\Phi_n(R_n)=\sum_zp_n(z)\tau_z^{(n)},\qquad
 b_n=\Phi_n(T_n)=\sum_zq_n(z)\tau_z^{(n)},
\]
the estimate \eqref{eq:channel-approx} bounds each of
$\|a_n-\rho\|_1$ and $\|b_n-\sigma\|_1$ by $2d^2/n$, and hence
\begin{equation}\label{eq:Bn-approx-pair}
 \|a_n-\rho\|_1+\|b_n-\sigma\|_1\le\frac{4d^2}{n}.
\end{equation}
These decompositions form a common classical reverse preparation of
$a_n,b_n$. Consequently, the reverse-test characterization
\eqref{eq:maximal-extension} and its specialization to the
Belavkin--Staszewski relative entropy give
\[
 \KL(p_n\Vert q_n)\ge\BS(a_n\Vert b_n).
\]

Since $\|X\|_\infty\leq\|X\|_1$ and
$\rho,\sigma\geq\lambda I$, the trace-distance
estimates imply
\[
a_n\geq
\left(\lambda-\frac{2d^2}{n}\right)I,
\qquad
b_n\geq
\left(\lambda-\frac{2d^2}{n}\right)I.
\]
Thus, for $n\geq n_0=\lceil4d^2/\lambda\rceil$,
both states are bounded below by $(\lambda/2)I$.

We apply the joint continuity estimate for BS relative
entropy of 
\cite[Theorem~10.6]{CapelCostaRico2026a}.
More precisely, for density operators $a,b,a',b'\ge(\lambda/2)I$,
the cited estimate with both lower eigenvalue bounds equal to
$\lambda/2$ gives
\begin{align}
 &\big|\BS(a\Vert b)-\BS(a'\Vert b')\big|\notag\\
 &\quad\le
 \frac12\left[\log\frac{4}{\lambda^2}
    +\left(\frac{2}{\lambda}\right)^{5/2}\right]\|a-a'\|_1
 +\log\left(1+\frac{\|b-b'\|_1}{\lambda}\right)\notag\\
 &\quad\le
 \left(\log\frac{2}{\lambda}
       +\frac{2\sqrt{2}}{\lambda^{5/2}}\right)\|a-a'\|_1
 +\frac{1}{\lambda}\|b-b'\|_1\notag\\
 &\quad\le L_{\rho,\sigma}
 \big(\|a-a'\|_1+\|b-b'\|_1\big),
 \label{eq:BS-explicit-Lipschitz}
\end{align}
where we used $\log(1+x)\le x$ and $0<\lambda\le1$.
Applying it to $(a_n,b_n)$ and
$(\rho,\sigma)$, and using \eqref{eq:Bn-approx-pair}, yields
\[
 \BS(a_n\Vert b_n)
 \ge\BS(\rho\Vert\sigma)
 -L_{\rho,\sigma}
 \big(\|a_n-\rho\|_1+\|b_n-\sigma\|_1\big)
 \ge\BS(\rho\Vert\sigma)-\frac{4d^2L_{\rho,\sigma}}{n}.
\]

Finally, the measurement channel associated with
$\{F_z^{(n)}\}_z$ and data processing give
\[
 \mathbb D(R_n\Vert T_n)
 \ge\mathbb D_{\mathrm{cl}}(p_n\Vert q_n)
 =\KL(p_n\Vert q_n)
 \ge\BS(\rho\Vert\sigma)-\frac{4d^2L_{\rho,\sigma}}{n}.
\]
Taking the infimum over compatible extensions using
Proposition~\ref{prop:parallel-preparation} yields the lower bound
in \eqref{eq:Bn-rate}. The upper bound follows from
\eqref{eq:horizon-general-hierarchy}, since the maximal extension
of classical relative entropy is $\BS$.
\end{proof}

\end{document}